\documentclass[12pt]{article}
\usepackage{a4wide,epsfig}

\usepackage{slashed}

\newcommand{\agt}{\rlap{\lower 3.5 pt \hbox{$\mathchar \sim$}} \raise 1pt
 \hbox {$>$}}
\newcommand{\alt}{\rlap{\lower 3.5 pt \hbox{$\mathchar \sim$}} \raise 1pt
 \hbox {$<$}}

\catcode`@=11
\newcount\@tempcntc
\def\@citex[#1]#2{\if@filesw\immediate\write\@auxout{\string\citation{#2}}\fi
  \@tempcnta\z@\@tempcntb\m@ne\def\@citea{}\@cite{\@for\@citeb:=#2\do
    {\@ifundefined
       {b@\@citeb}{\@citeo\@tempcntb\m@ne\@citea\def\@citea{,}{\bf
?}\@warning
       {Citation `\@citeb' on page \thepage \space undefined}}%
    {\setbox\z@\hbox{\global\@tempcntc0\csname b@\@citeb\endcsname\relax}%
     \ifnum\@tempcntc=\z@ \@citeo\@tempcntb\m@ne
       \@citea\def\@citea{,}\hbox{\csname b@\@citeb\endcsname}%
     \else
      \advance\@tempcntb\@ne
      \ifnum\@tempcntb=\@tempcntc
      \else\advance\@tempcntb\m@ne\@citeo
      \@tempcnta\@tempcntc\@tempcntb\@tempcntc\fi\fi}}\@citeo}{#1}}
\def\@citeo{\ifnum\@tempcnta>\@tempcntb\else\@citea\def\@citea{,}%
  \ifnum\@tempcnta=\@tempcntb\the\@tempcnta\else
   {\advance\@tempcnta\@ne\ifnum\@tempcnta=\@tempcntb \else
\def\@citea{--}\fi
    \advance\@tempcnta\m@ne\the\@tempcnta\@citea\the\@tempcntb}\fi\fi}
\catcode`@=12

\begin{document}

\title{
\vskip-3cm{\baselineskip14pt
\centerline{\normalsize DESY 11--046\hfill ISSN 0418-9833}
\centerline{\normalsize May 2011\hfill}}
\vskip1.5cm
World data of $J/\psi$ production consolidate NRQCD factorization at NLO}

\author{Mathias Butensch\"on, Bernd A. Kniehl\\
{\normalsize II. Institut f\"ur Theoretische Physik, Universit\"at Hamburg,}\\
{\normalsize Luruper Chaussee 149, 22761 Hamburg, Germany}
}
\date{}

\maketitle

\begin{abstract}
We calculate the cross sections of inclusive $J/\psi$ production in
photoproduction and two-photon scattering, involving both direct and resolved
photons, and in $e^+e^-$ annihilation at next-to-leading order (NLO) within the
factorization formalism of nonrelativistic quantum chromodynamics (NRQCD),
including the full relativistic corrections due to the intermediate
$^1\!S_0^{[8]}$, $^3\!S_1^{[8]}$, and $^3\!P_J^{[8]}$ color-octet (CO) states.
Exploiting also our previous results on hadroproduction, we perform a
combined fit of the respective CO long-distance matrix elements (LDMEs) to all
available high-quality data of inclusive $J/\psi$ production, from KEKB,
LEP~II, RHIC, HERA, the Tevatron, and the LHC, comprising a total of 194 data
points from 26 data sets.

\medskip

\noindent
PACS numbers: 12.38.Bx, 13.60.Le, 13.85.Ni, 14.40.Gx
\end{abstract}

\newpage

The factorization formalism of NRQCD \cite{Bodwin:1994jh}
provides a rigorous theoretical framework for the description of
heavy-quarkonium production and decay.
This implies a separation of process-dependent short-distance coefficients, to
be calculated perturbatively as expansions in the strong-coupling constant
$\alpha_s$, from supposedly universal LDMEs, to be extracted from experiment.
The relative importance of the latter can be estimated by means of velocity
scaling rules; {\it i.e.}, the LDMEs are predicted to scale with a definite
power of the heavy-quark ($Q$) velocity $v$ in the limit $v\ll1$.
In this way, the theoretical predictions are organized as double expansions in
$\alpha_s$ and $v$.
A crucial feature of this formalism is that it takes into account the complete
structure of the $Q\overline{Q}$ Fock space, which is spanned by the states
$n={}^{2S+1}L_J^{[a]}$ with definite spin $S$, orbital angular momentum
$L$, total angular momentum $J$, and color multiplicity $a=1,8$.
In particular, this formalism predicts the existence of CO processes in nature.
This means that $Q\overline{Q}$ pairs are produced at short distances in
CO states and subsequently evolve into physical, color-singlet (CS) quarkonia
by the nonperturbative emission of soft gluons.
In the limit $v\to0$, the traditional CS model (CSM) is recovered in the case
of $S$-wave quarkonia.
In the case of $J/\psi$ production, the CSM prediction is based just on the
$^3\!S_1^{[1]}$ CS state, while the leading relativistic corrections, of
relative order ${\cal O}(v^4)$, are built up by the $^1\!S_0^{[8]}$,
$^3\!S_1^{[8]}$, and $^3\!P_J^{[8]}$ ($J=0,1,2$) CO states.
The CSM is not a complete theory, as may be understood by noticing that the NLO
treatment of $P$-wave quarkonia is plagued by uncanceled infrared
singularities, which are, however, properly removed in NRQCD.

The test of NRQCD factorization has been identified to be among the most
exigent milestones on the roadmap of quarkonium physics at the present time
\cite{Brambilla:2010cs}. 
While, for $J/\psi$ polarization, comparisons of HERA and Tevatron data with
NRQCD predictions, which are not yet fully known at NLO, unravel a rather
confusing pattern, the situation is eventually clearing up for the $J/\psi$
yield, which is now fully known at NLO in NRQCD for direct photoproduction
\cite{Butenschoen:2009zy} and hadroproduction
\cite{Butenschoen:2010rq,Ma:2010yw}.
In fact, it has been demonstrated \cite{Butenschoen:2010rq} that the set of CO
LDMEs fitted to transverse-momentum ($p_T$) distributions measured at HERA
\cite{Adloff:2002ex,Aaron:2010gz} and by CDF at Tevatron~II
\cite{Acosta:2004yw} also lead to very good descriptions of distributions in
the $\gamma p$ c.m.\ energy $W$ and the inelasticity $z$, which measures the
fraction of $\gamma$ energy passed on to the $J/\psi$ meson in the $p$ rest
frame, from HERA \cite{Adloff:2002ex,Aaron:2010gz} and of $p_T$ distributions
from RHIC \cite{Adare:2009js} and the LHC \cite{Khachatryan:2010yr}.
On the other hand, the Tevatron~II \cite{Acosta:2004yw} data alone can only pin
down two linear combinations of the three CO LDMEs
\cite{Ma:2010yw,Butenschoen:2010px}, and the fit results of
Ref.~\cite{Ma:2010yw} are incompatible with Ref.~\cite{Butenschoen:2010rq}.
It is the purpose of this Letter, to overcome this highly unsatisfactory
situation jeopardizing the success of NRQCD factorization by performing a
global fit to all available high-quality data of inclusive unpolarized $J/\psi$
production, comprising a total of 194 data points from 26 data sets.
Specifically, these include $p_T$ distributions in hadroproduction from PHENIX
\cite{Adare:2009js} at RHIC, CDF at Tevatron~I \cite{Abe:1997jz} and
Tevatron~II \cite{Acosta:2004yw}, ATLAS \cite{ATLASdata}, CMS
\cite{Khachatryan:2010yr}, ALICE \cite{ALICEdata}, and LHCb \cite{LHCbdata} at
the LHC; $p_T^2$, $W$, and $z$ distributions in photoproduction from ZEUS
\cite{Chekanov:2002at} and H1 \cite{Adloff:2002ex} at HERA~I and H1
\cite{Aaron:2010gz} at HERA~II; a $p_T^2$ distribution in two-photon scattering
from DELPHI \cite{Abdallah:2003du} at LEP~II; and a total cross section in
$e^+e^-$ annihilation from Belle \cite{:2009nj} at KEKB.

Incoming photons participate in the hard scattering either directly or via
partons into which they fluctuate (resolve) intermittently, and both modes of
interaction contribute at the same order of perturbation theory.
Therefore, we need to extend the theoretical ingredients available from
Refs.~\cite{Butenschoen:2009zy,Butenschoen:2010rq} by also treating
$\gamma p\to J/\psi+X$ with the photon being resolved and
$\gamma\gamma\to J/\psi+X$ with none \cite{Klasen:2004tz}, one, or both of the
photons being resolved at NLO in NRQCD.
We repeat the analysis of Ref.~\cite{Klasen:2004tz}, in which the Coulomb
singularities were regularized by $v$, using dimensional regularization as in
Refs.~\cite{Butenschoen:2009zy,Butenschoen:2010rq} in order to obtain analytic
expressions sufficiently compact for our purposes.
We also find it necessary to revisit $e^+e^-\to J/\psi+X$ at NLO in NRQCD
because the results of Ref.~\cite{Zhang:2009ym} have not yet been verified by
an independent calculation, are only available in numerical form, and lack the
$^3\!S_1^{[8]}$ contribution, which comes both with $X=q\overline{q}$
\cite{Yuan:1996ep} and $gg$.
Higher-order corrections to the CSM process
$e^+e^-\to c\overline{c}[^3\!S_1^{[1]}]gg$ \cite{Ma:2008gq}, which enters our
analysis at NLO, are beyond the order considered here. 

The additional analytic calculations proceed along the lines of
Refs.~\cite{Butenschoen:2009zy,Butenschoen:2010rq} and are not described here
in detail for lack of space.
We merely present our master formula based on the factorization theorems of
the QCD parton model and NRQCD \cite{Bodwin:1994jh}:
\begin{eqnarray}
d\sigma(AB\to J/\psi+X)
&=&\sum_{i,j,k,l,n} \int dx_1dx_2dy_1dy_2\,f_{i/A}(x_1)
f_{k/i}(y_1)f_{j/B}(x_2)f_{l/j}(y_2)
\nonumber\\
&&{}\times \langle{\cal O}^{J/\psi}[n]\rangle
d\sigma(k l\to c\overline{c}[n]+X),
\label{Overview.Cross}
\end{eqnarray}
where $f_{i/A}(x_1)$ is the parton distribution function (PDF) of parton
$i=g,q,\overline{q}$ in hadron $A=p,\overline{p}$ or the flux function of
photon $i=\gamma$ in charged lepton $A=e^-,e^+$, $f_{k/i}(y_1)$ is
$\delta_{ik}\delta(1-y_1)$ or the PDF of parton $k$ in the resolved photon $i$,
$d\sigma(k l\to c\overline{c}[n]+X)$ are the partonic cross sections, and
$\langle{\cal O}^{J/\psi}[n]\rangle$ are the LDMEs.
In the fixed-flavor-number scheme, we have $q=u,d,s$.
In the case of $e^+e^-$ annihilation, all distribution functions in
Eq.~(\ref{Overview.Cross}) are delta functions.
As in Refs.~\cite{Butenschoen:2009zy,Butenschoen:2010rq}, $X$ always contains
one hard parton at leading order (LO) and is void of heavy flavors, which may
be tagged and vetoed experimentally.

\begin{table}
\begin{center}
\begin{tabular}{|c|c|}
\hline
$\langle {\cal O}^{J/\psi}(^1\!S_0^{[8]}) \rangle$ &
$(4.97\pm0.44)\times10^{-2}$~GeV$^3$ \\
$\langle {\cal O}^{J/\psi}(^3\!S_1^{[8]}) \rangle$ &
$(2.24\pm0.59)\times10^{-3}$~GeV$^3$ \\
$\langle {\cal O}^{J/\psi}(^3\!P_0^{[8]}) \rangle$ &
$(-1.61\pm0.20)\times10^{-2}$~GeV$^5$ \\
\hline
\end{tabular}
\end{center}
\caption{\label{tab:fit} NLO fit results for the $J/\psi$ CO LDMEs.}
\end{table}
We now describe our theoretical input for our numerical analyses.
We set $m_c=1.5$~GeV, adopt the values of $m_e$, $\alpha$, and the branching
ratios  $B(J/\psi\to e^+e^-)$ and $B(J/\psi\to\mu^+\mu^-)$ from
Ref.~\cite{Nakamura:2010zzi}, and use the one-loop (two-loop) formula for
$\alpha_s^{(n_f)}(\mu)$, with $n_f=4$ active quark flavors, at LO (NLO).
As for the proton PDFs, we use set CTEQ6L1 (CTEQ6M) \cite{Pumplin:2002vw} at
LO (NLO), which comes with an asymptotic scale parameter of
$\Lambda_\mathrm{QCD}^{(4)}=215$~MeV (326~MeV).
As for the photon PDFs, we employ the best-fit set AFG04\_BF of
Ref.~\cite{Aurenche:2005da}.
We evaluate the photon flux function using Eq.~(5) of
Ref.~\cite{Kniehl:1996we}, with the upper cutoff on the photon virtuality $Q^2$
chosen as in the considered data set.
As for the CS LDME, we adopt the value
$\langle {\cal O}^{J/\psi}(^3\!S_1^{[1]}) \rangle = 1.32$~GeV$^3$ from
Ref.~\cite{Bodwin:2007fz}.
Our default choices for the renormalization, factorization, and NRQCD scales
are $\mu_r=\mu_f=m_T$ and $\mu_\Lambda=m_c$, respectively, where
$m_T=\sqrt{p_T^2+4m_c^2}$ is the $J/\psi$ transverse mass.
The bulk of the theoretical uncertainty is due to the lack of knowledge of
corrections beyond NLO, which are estimated by varying $\mu_r$, $\mu_f$, and
$\mu_\Lambda$ by a factor 2 up and down relative to their default values.

\begin{figure*}
\includegraphics[width=0.225\textwidth]{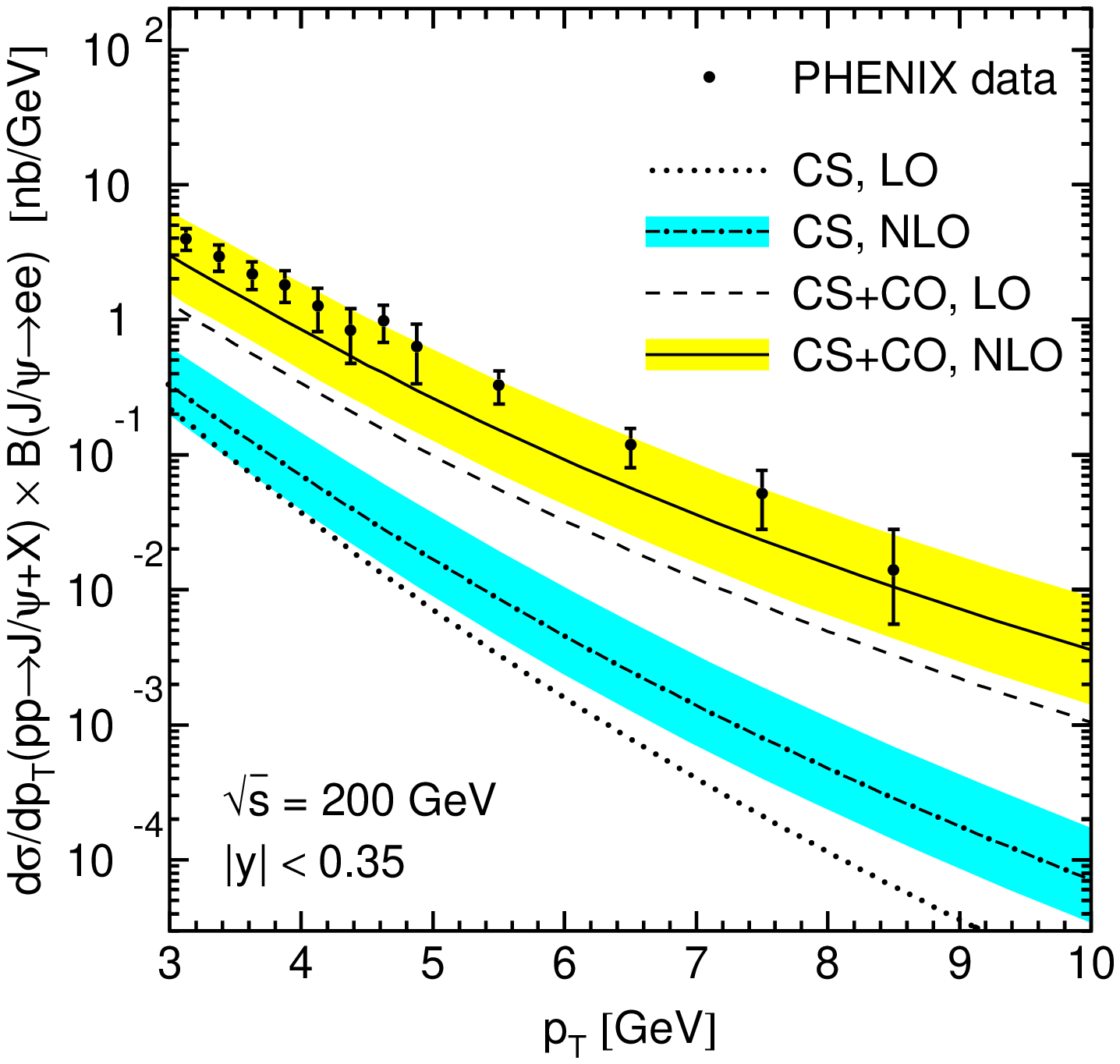}
\includegraphics[width=0.225\textwidth]{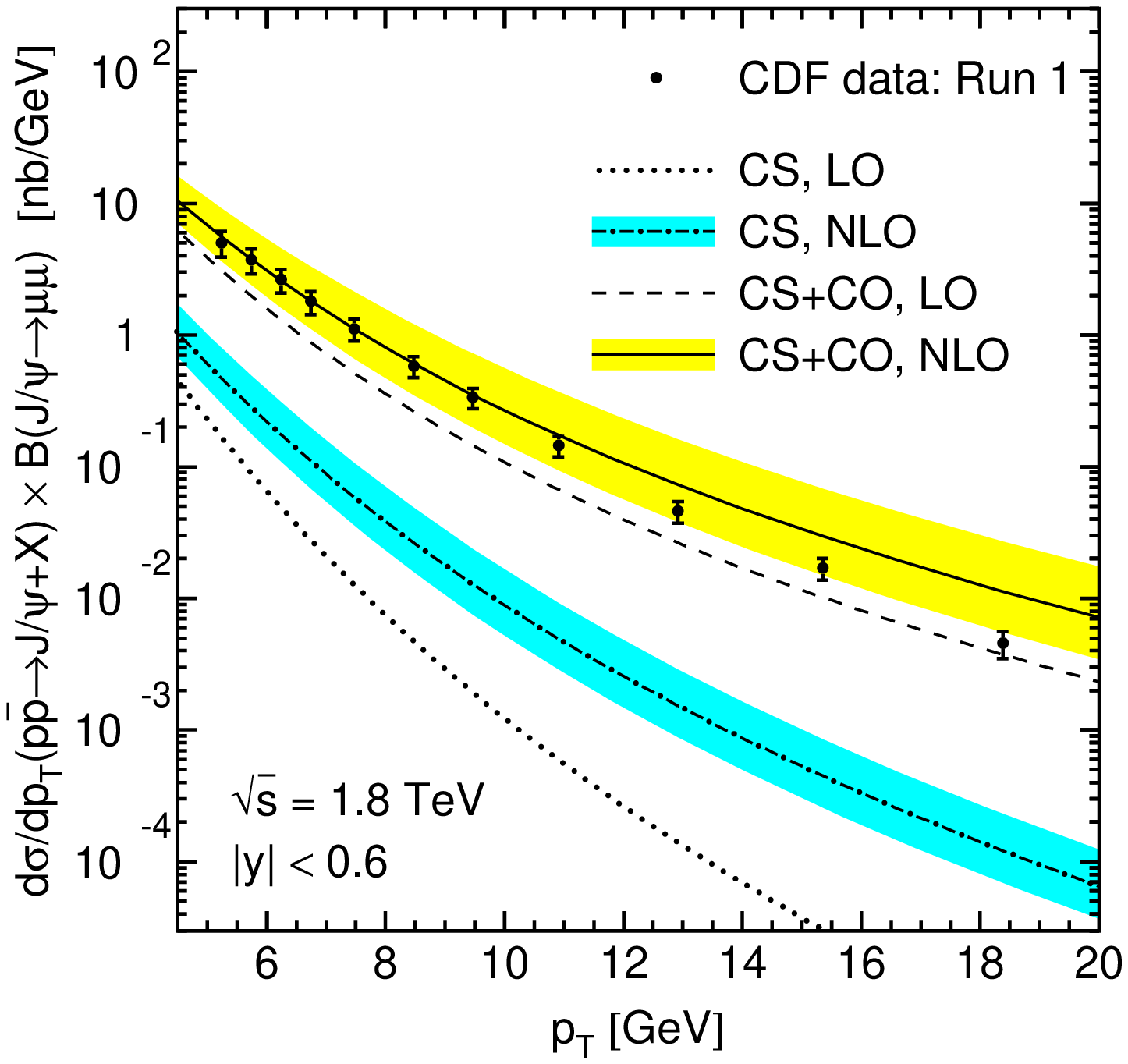}
\includegraphics[width=0.225\textwidth]{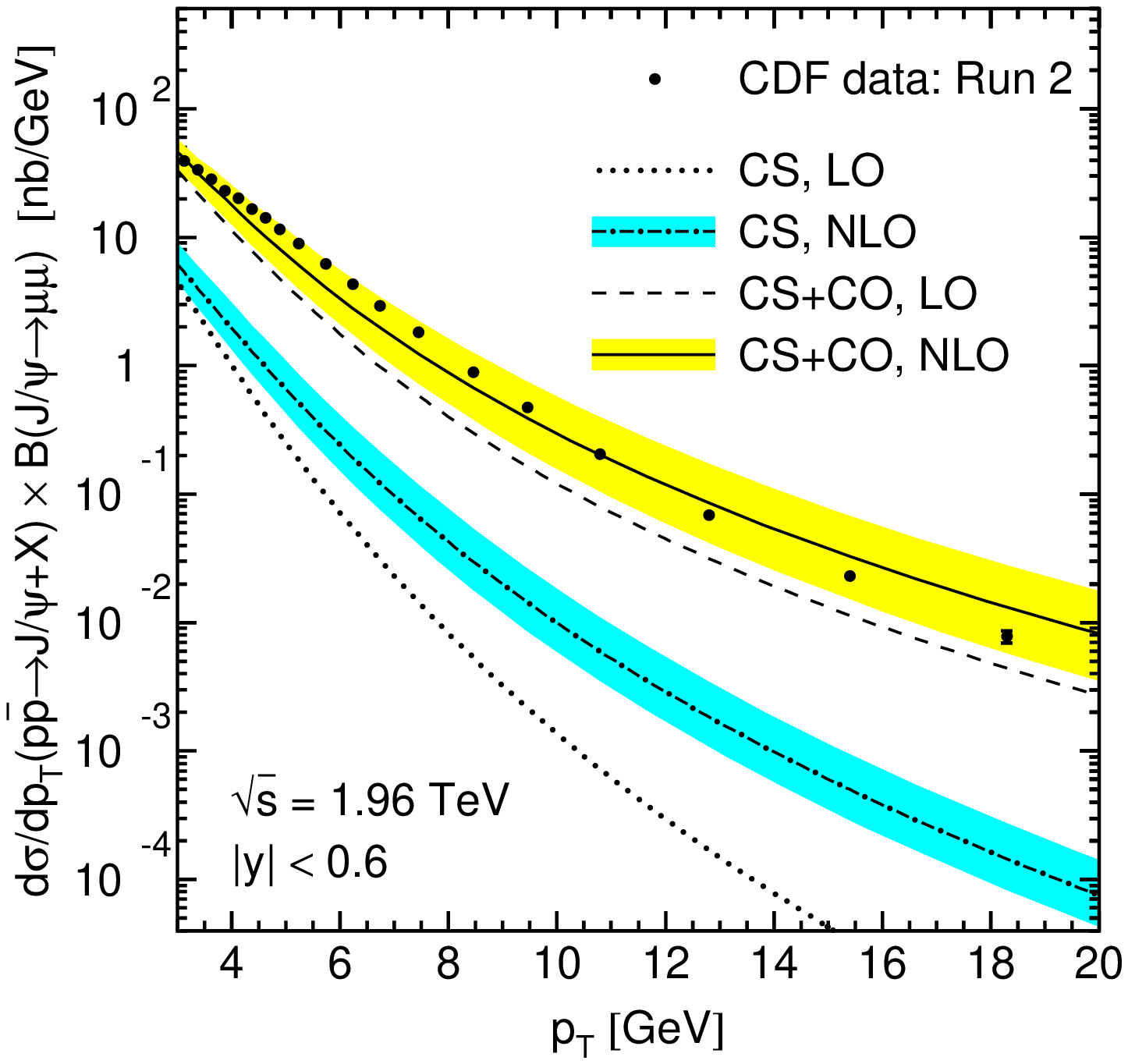}
\includegraphics[width=0.225\textwidth]{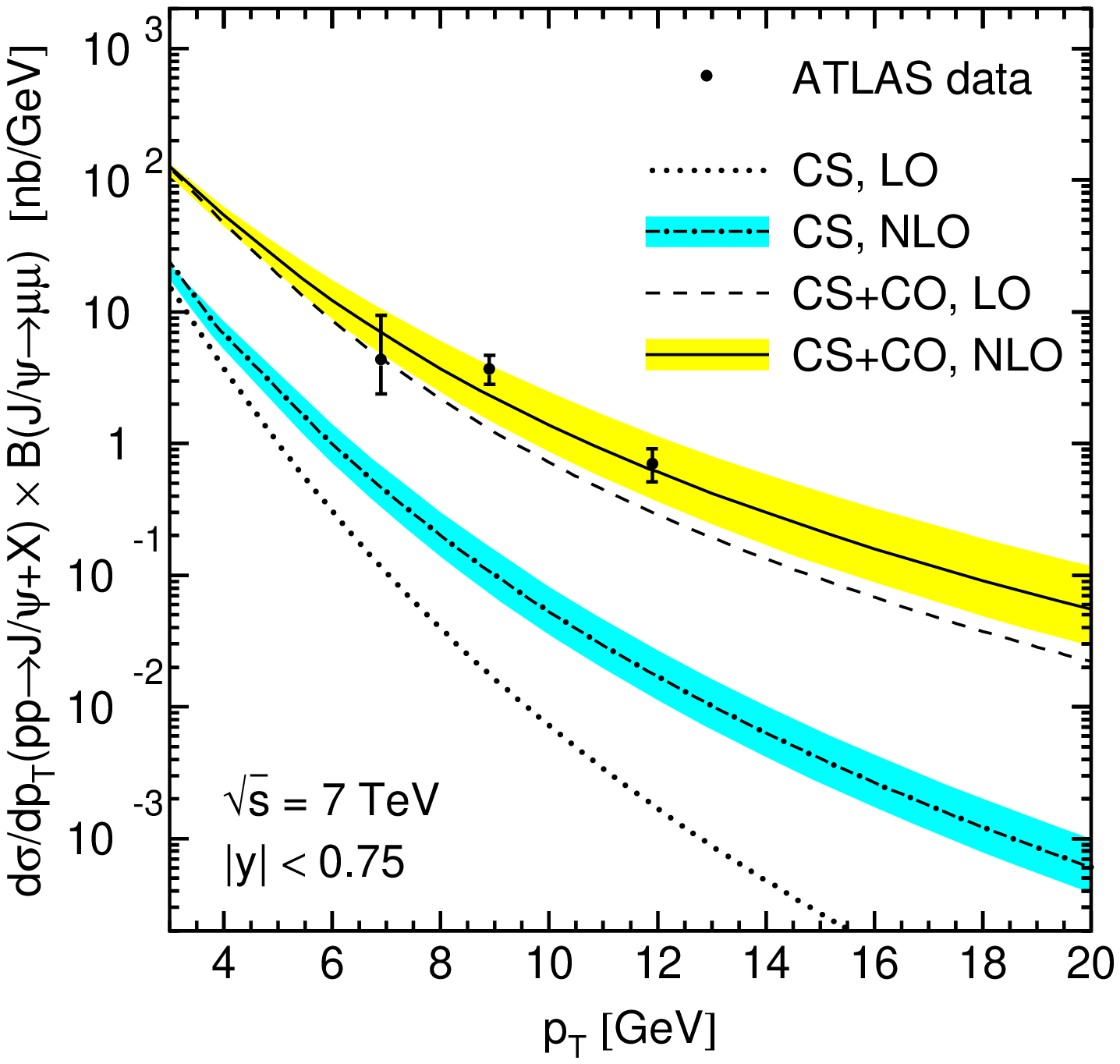}

\vspace{2pt}
\includegraphics[width=0.225\textwidth]{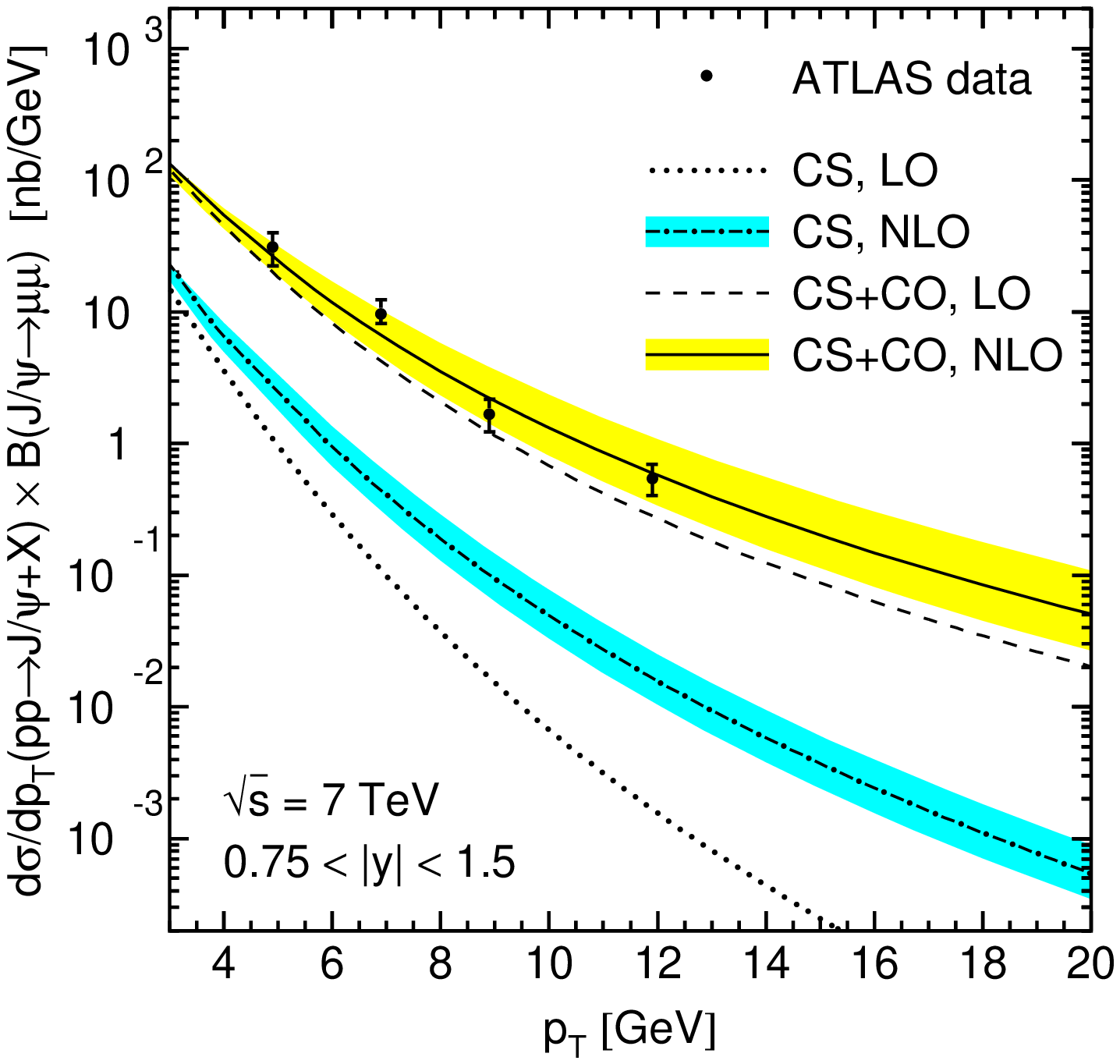}
\includegraphics[width=0.225\textwidth]{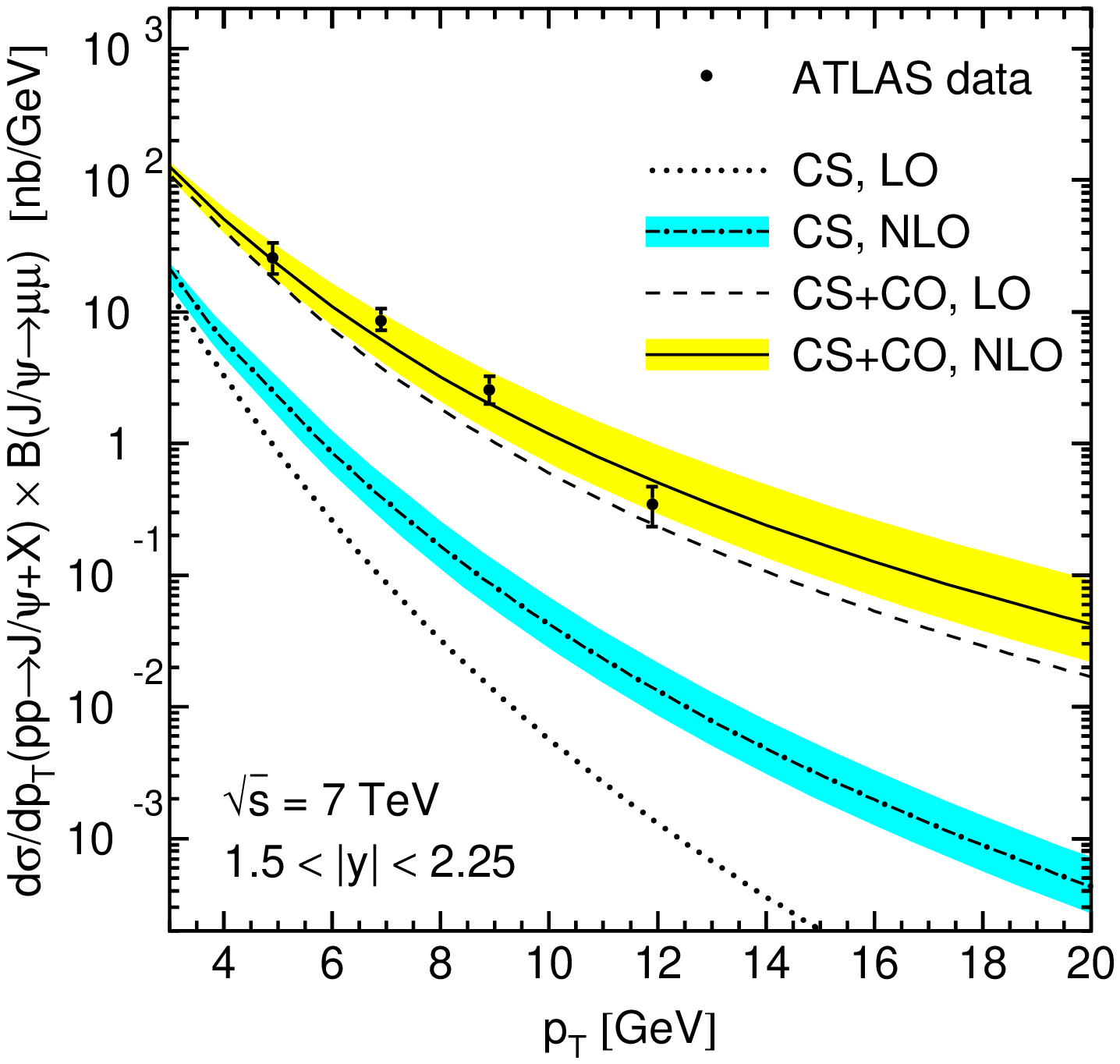}
\includegraphics[width=0.225\textwidth]{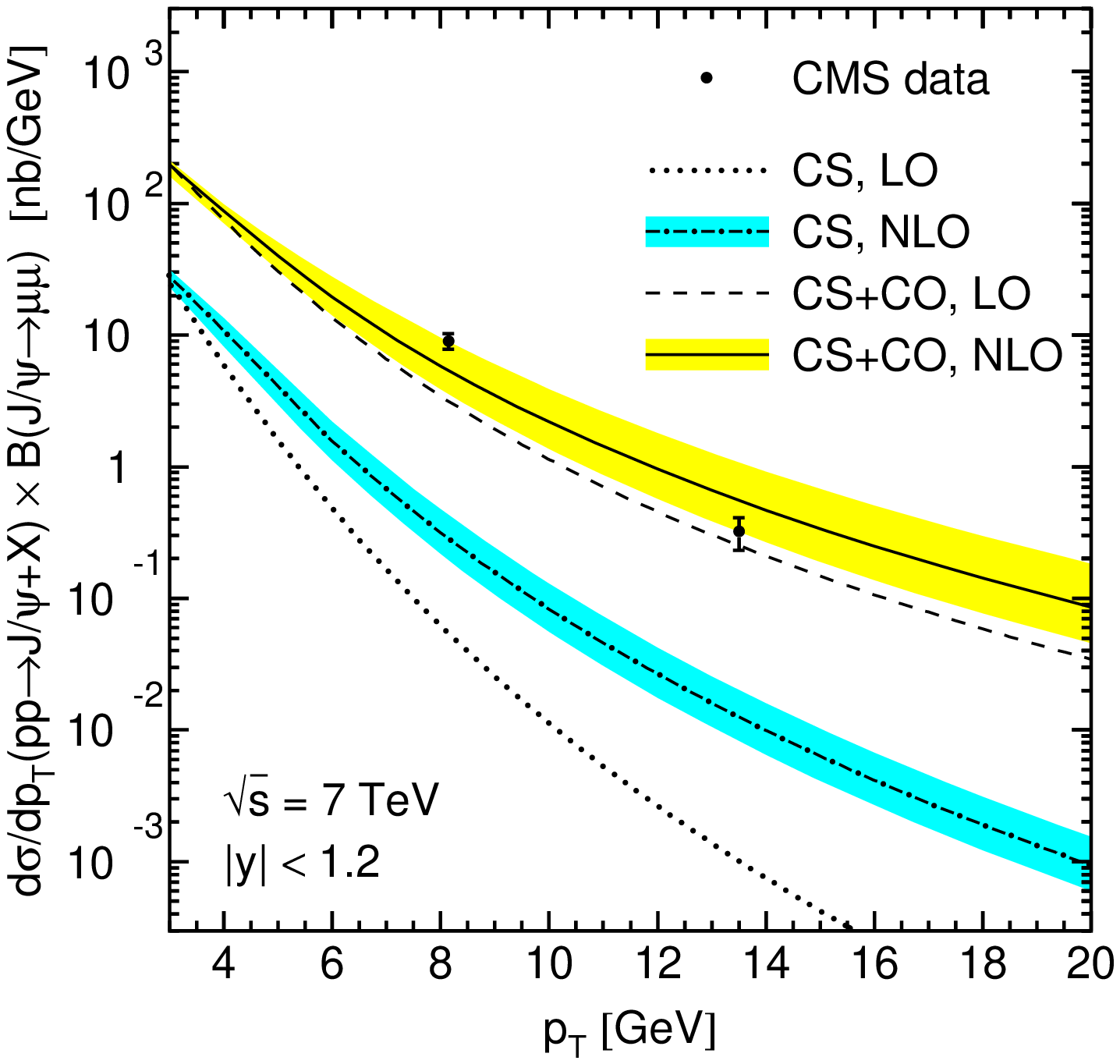}
\includegraphics[width=0.225\textwidth]{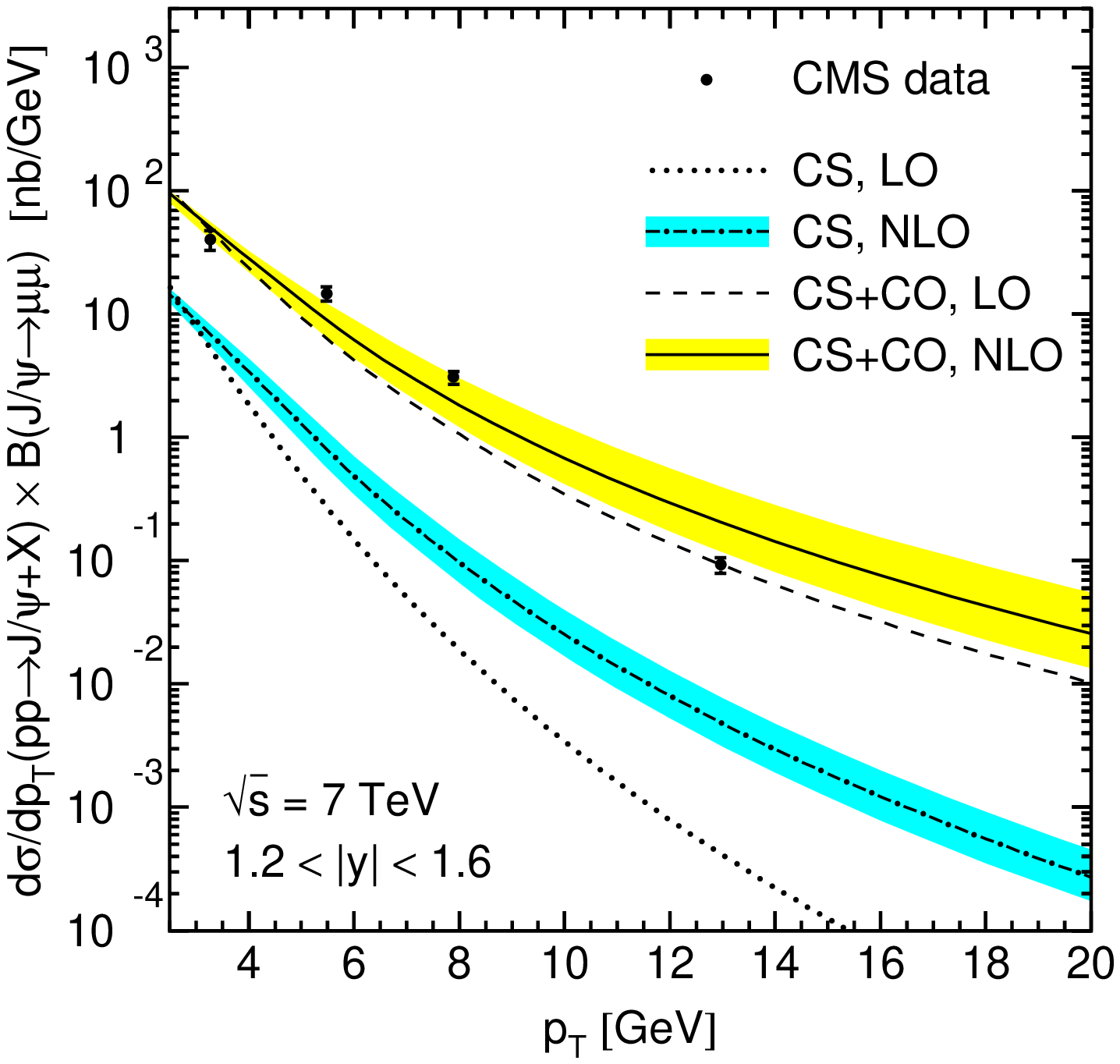}

\vspace{2pt}
\includegraphics[width=0.225\textwidth]{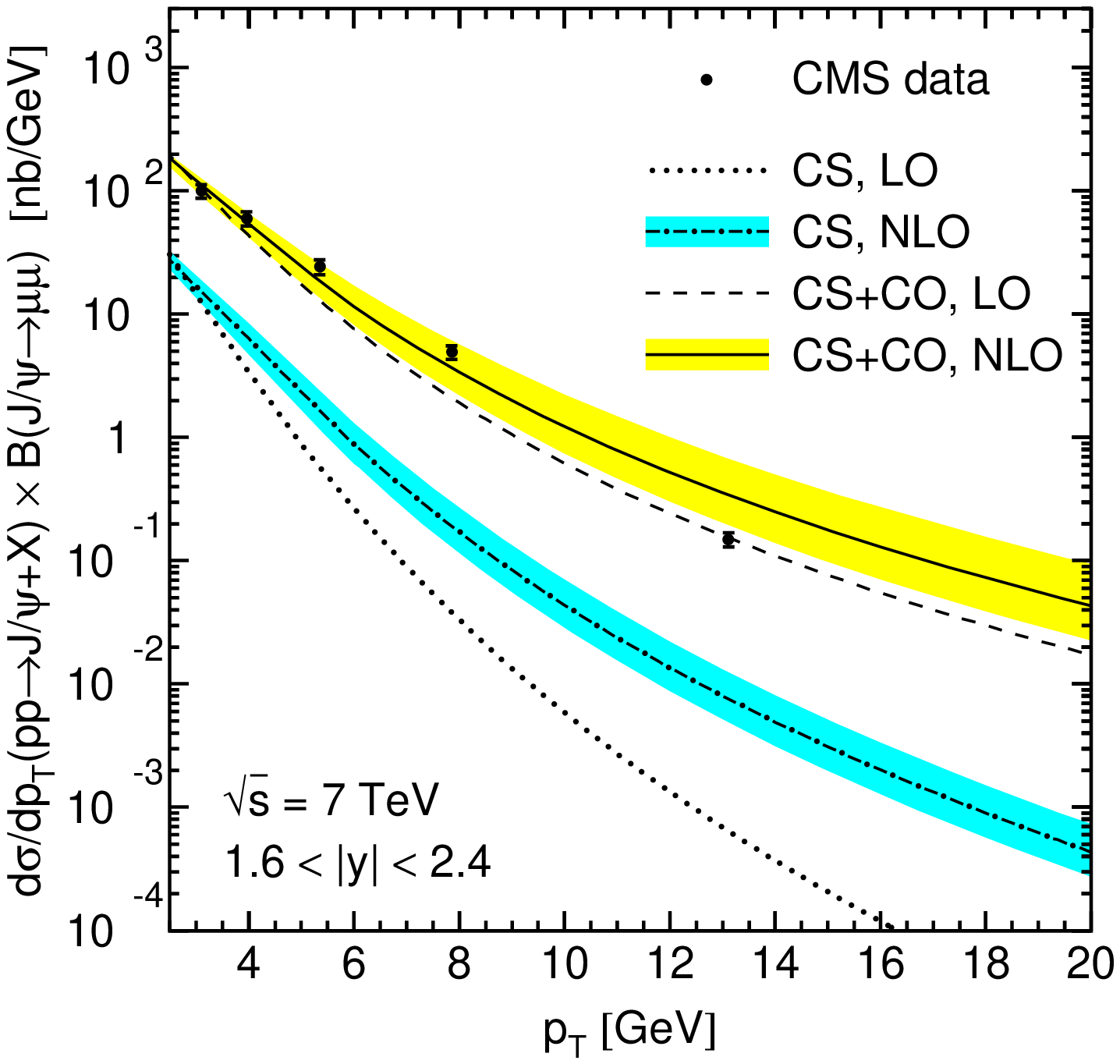}
\includegraphics[width=0.225\textwidth]{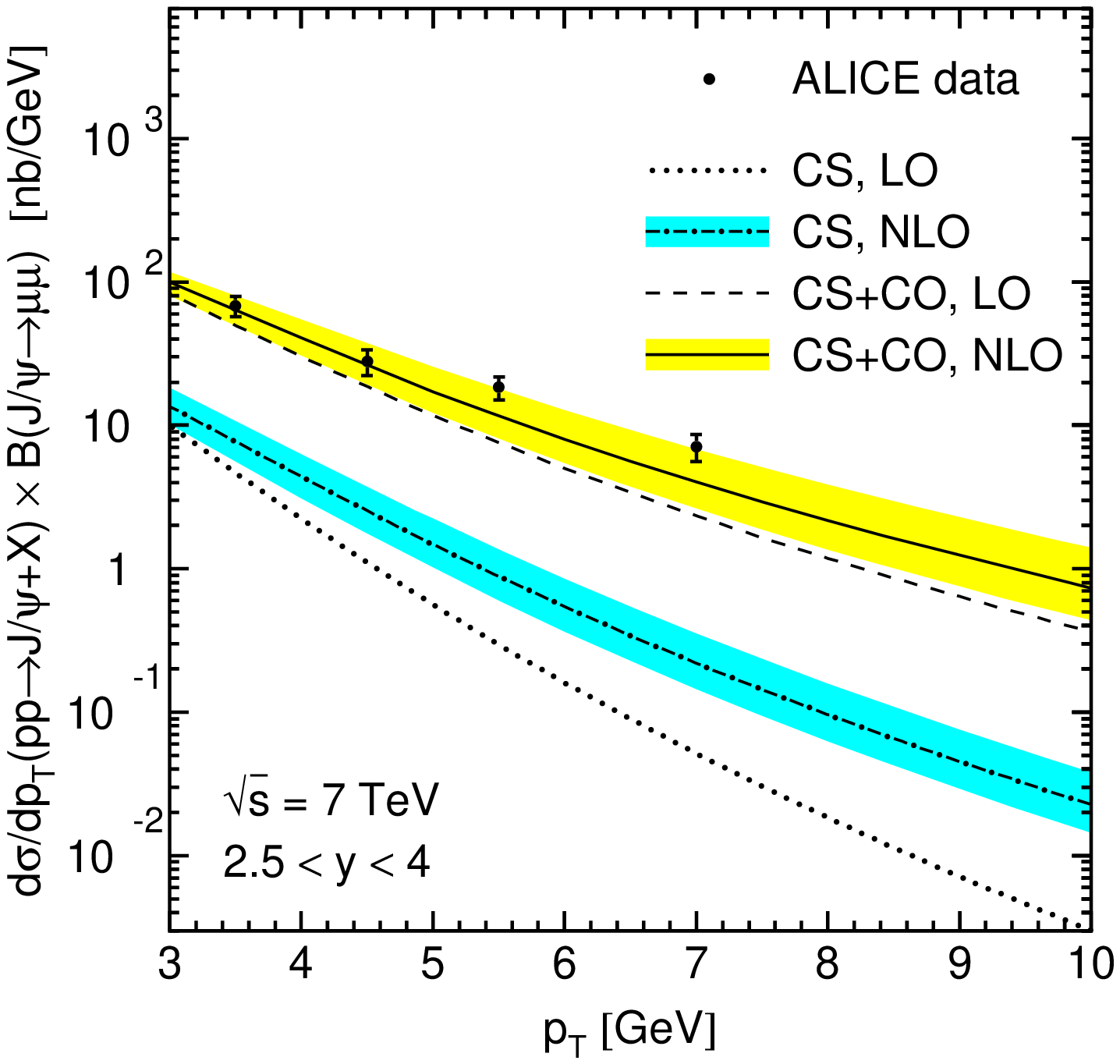}
\includegraphics[width=0.225\textwidth]{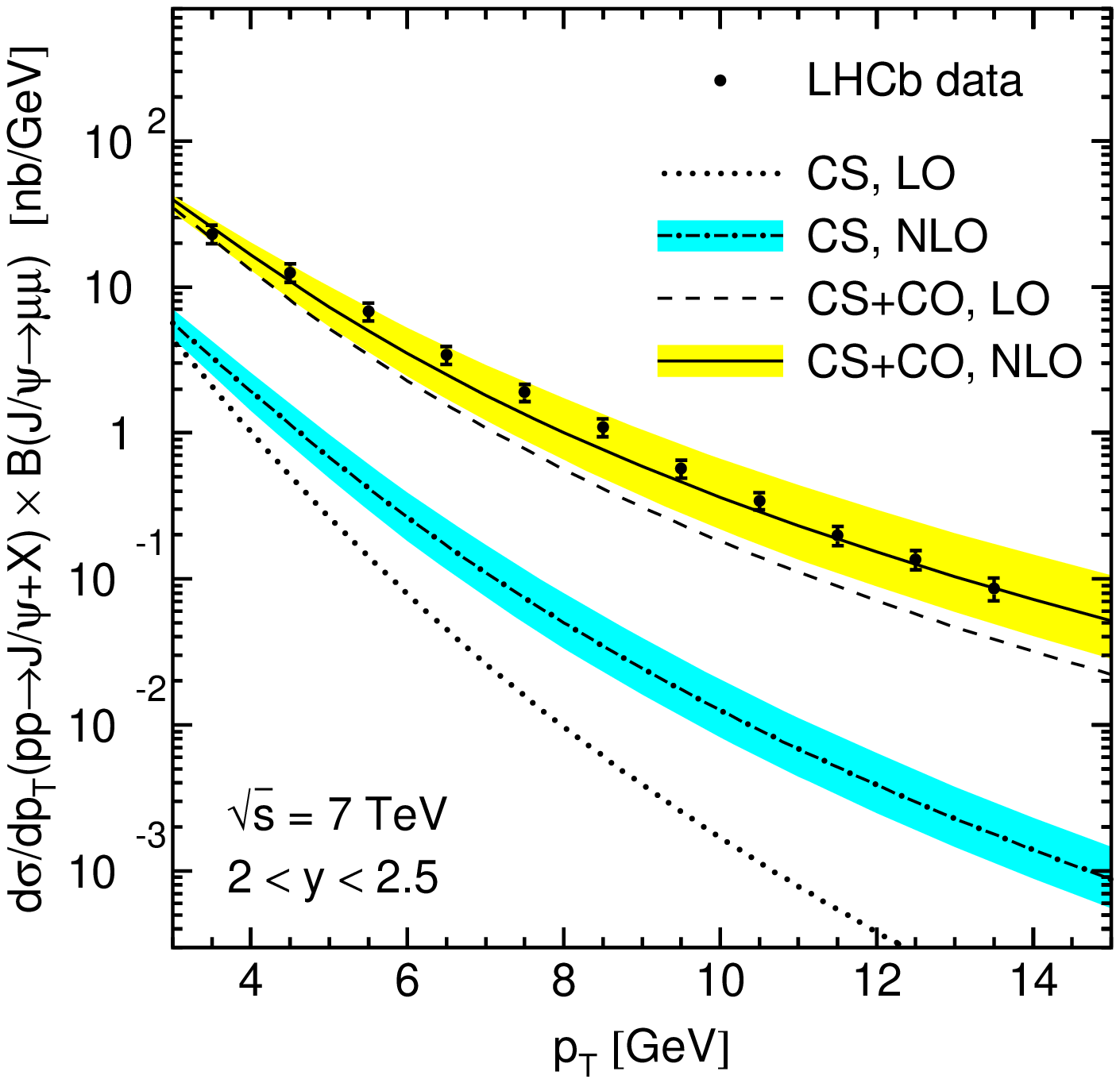}
\includegraphics[width=0.225\textwidth]{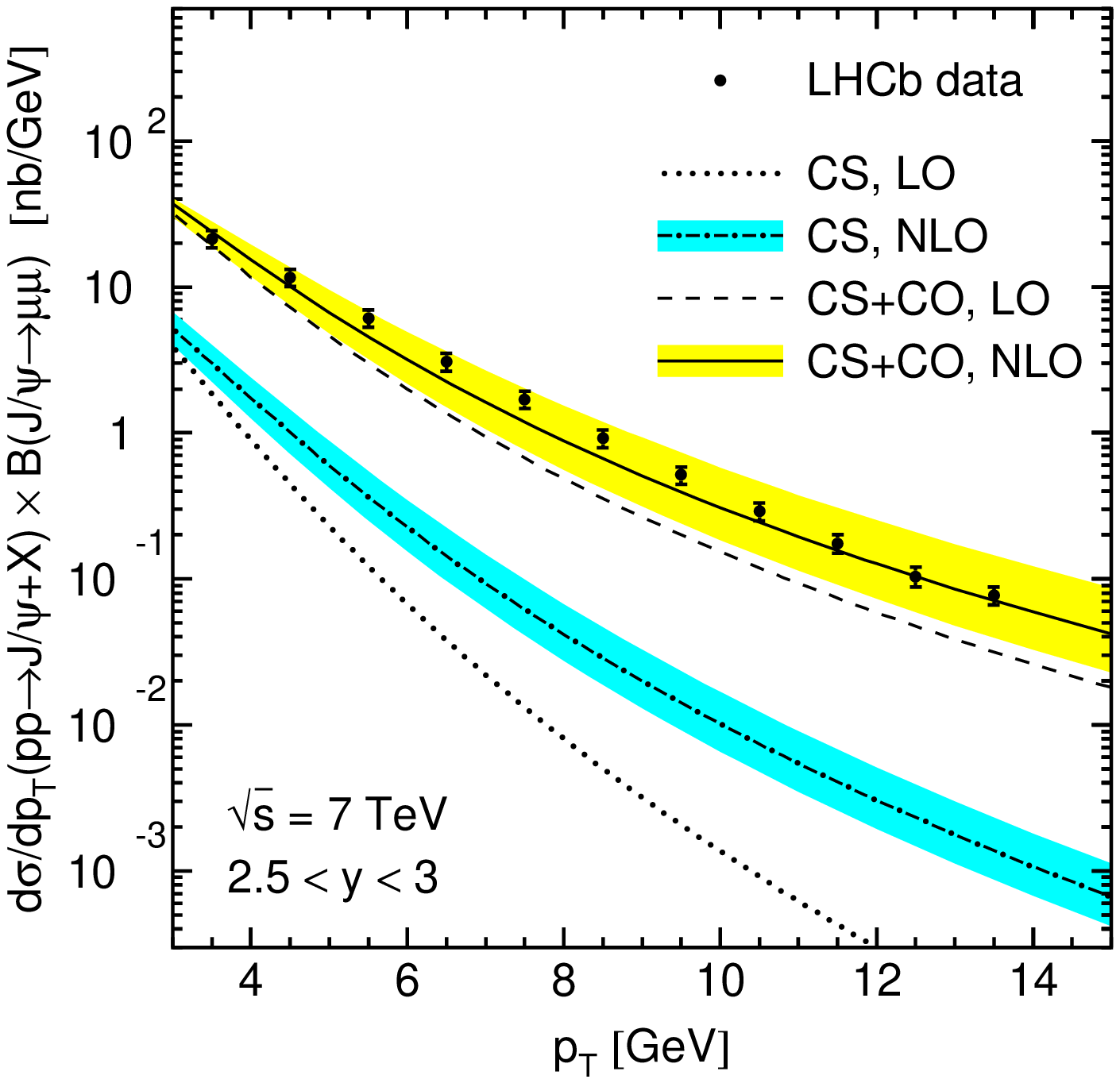}

\vspace{2pt}
\includegraphics[width=0.225\textwidth]{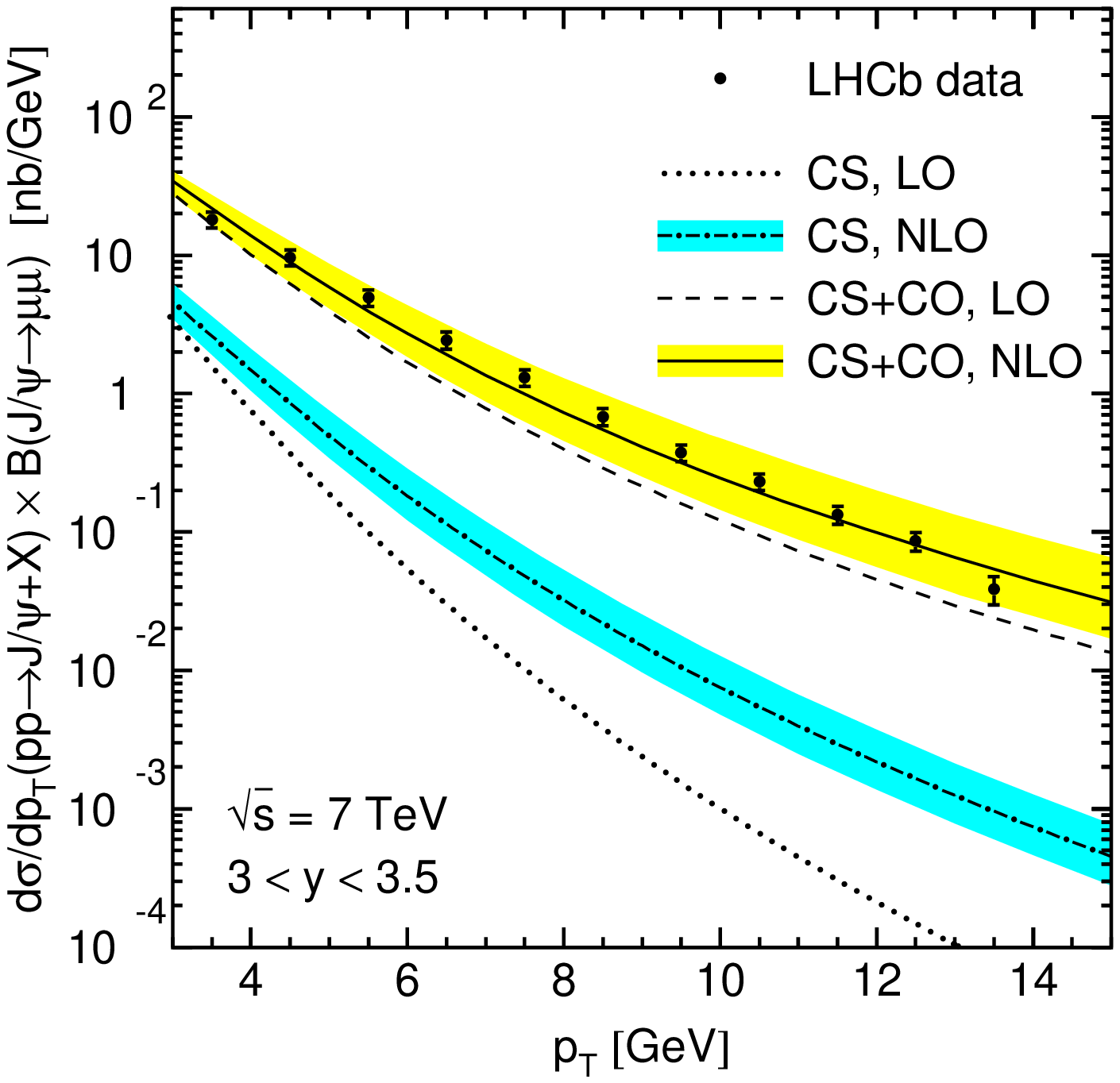}
\includegraphics[width=0.225\textwidth]{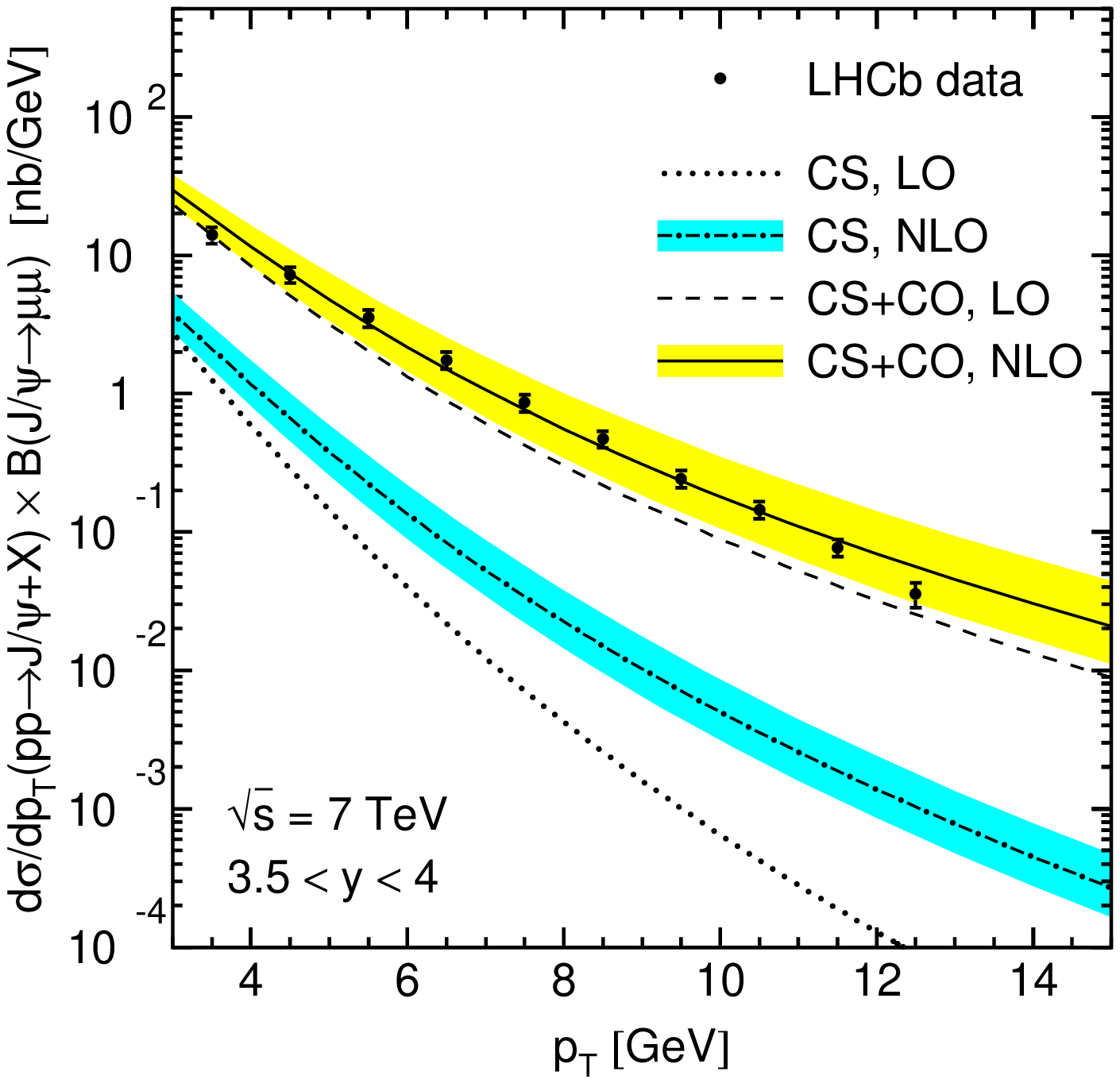}
\includegraphics[width=0.225\textwidth]{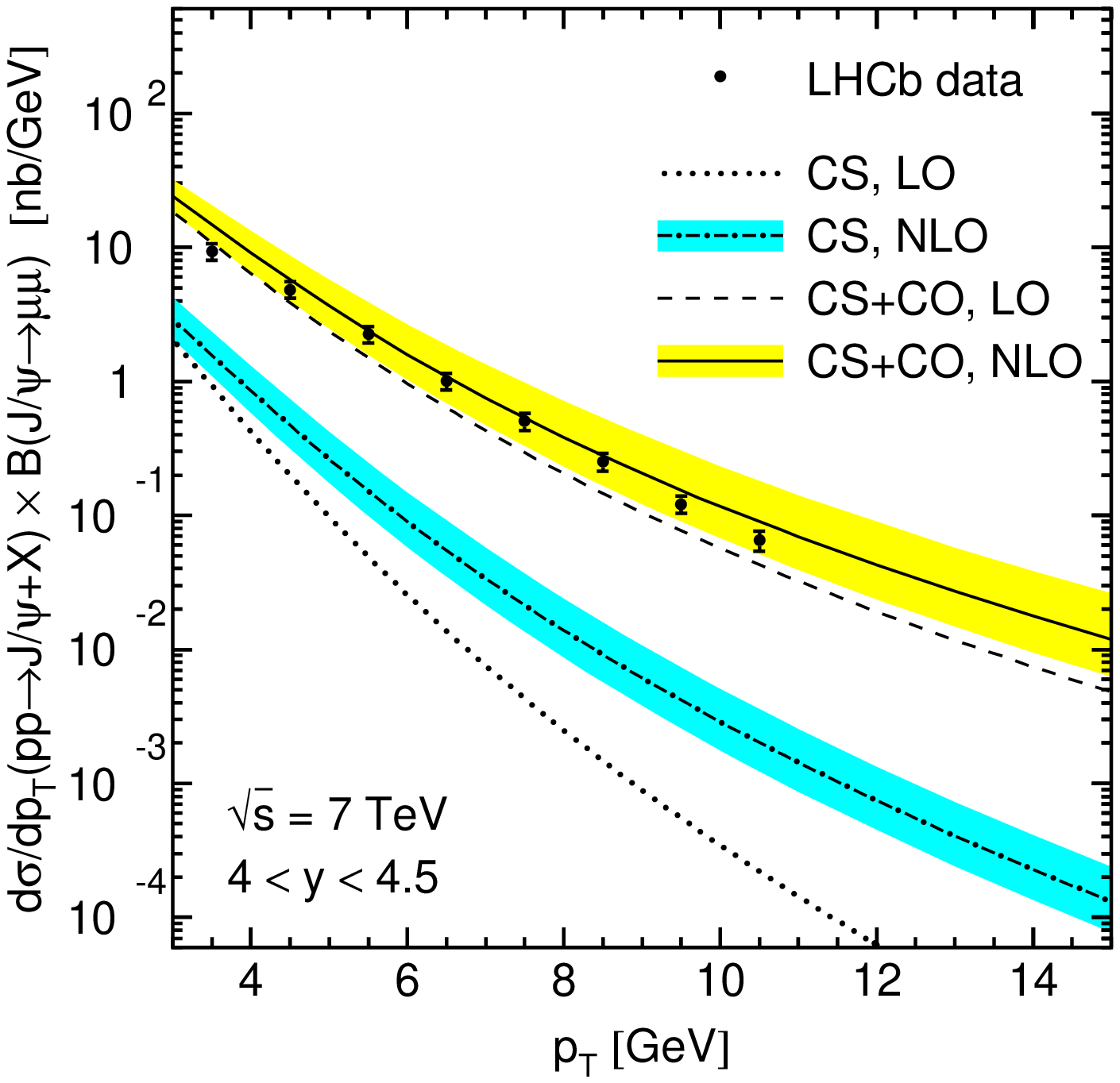}
\includegraphics[width=0.225\textwidth]{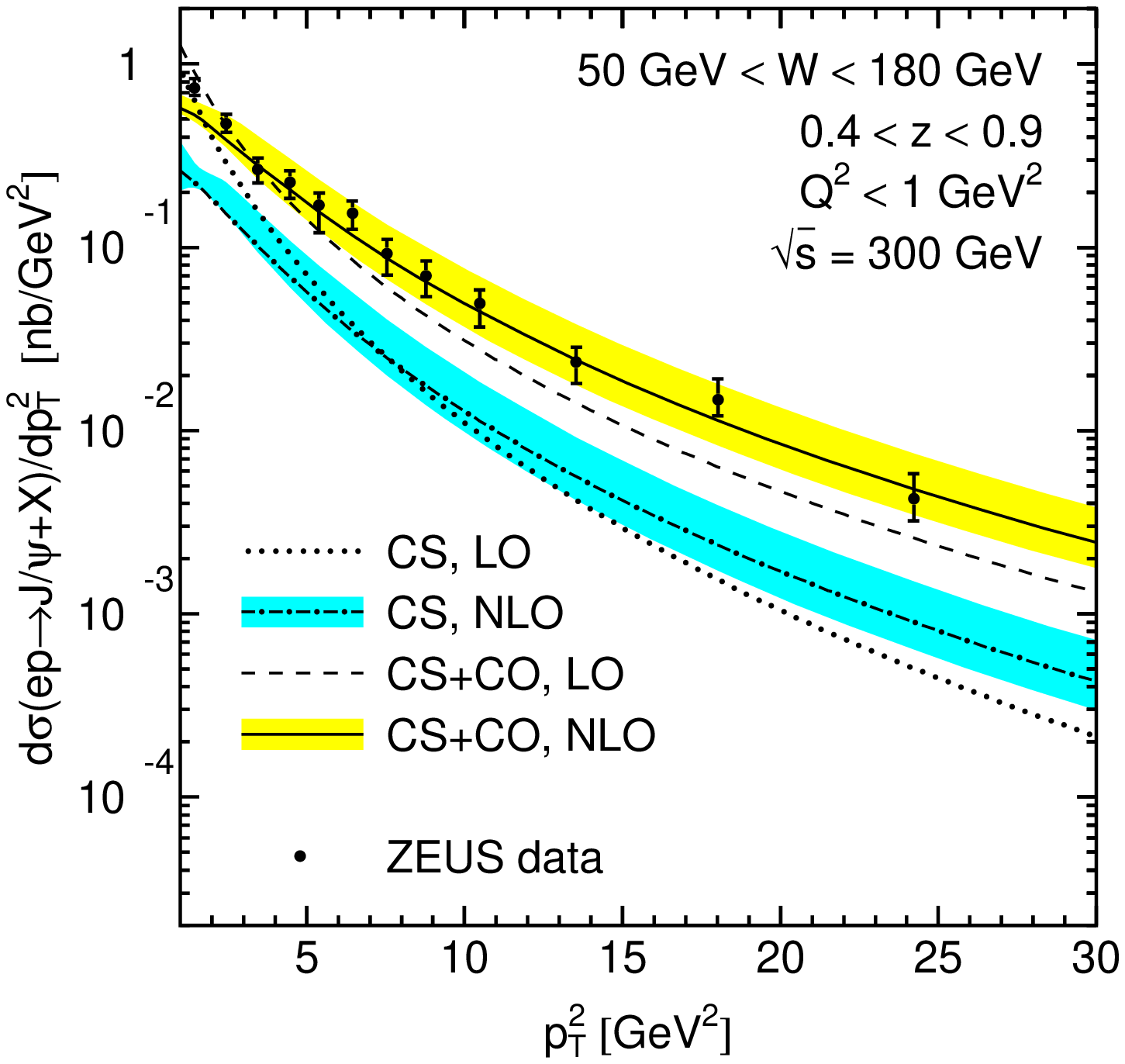}

\vspace{2pt}
\includegraphics[width=0.225\textwidth]{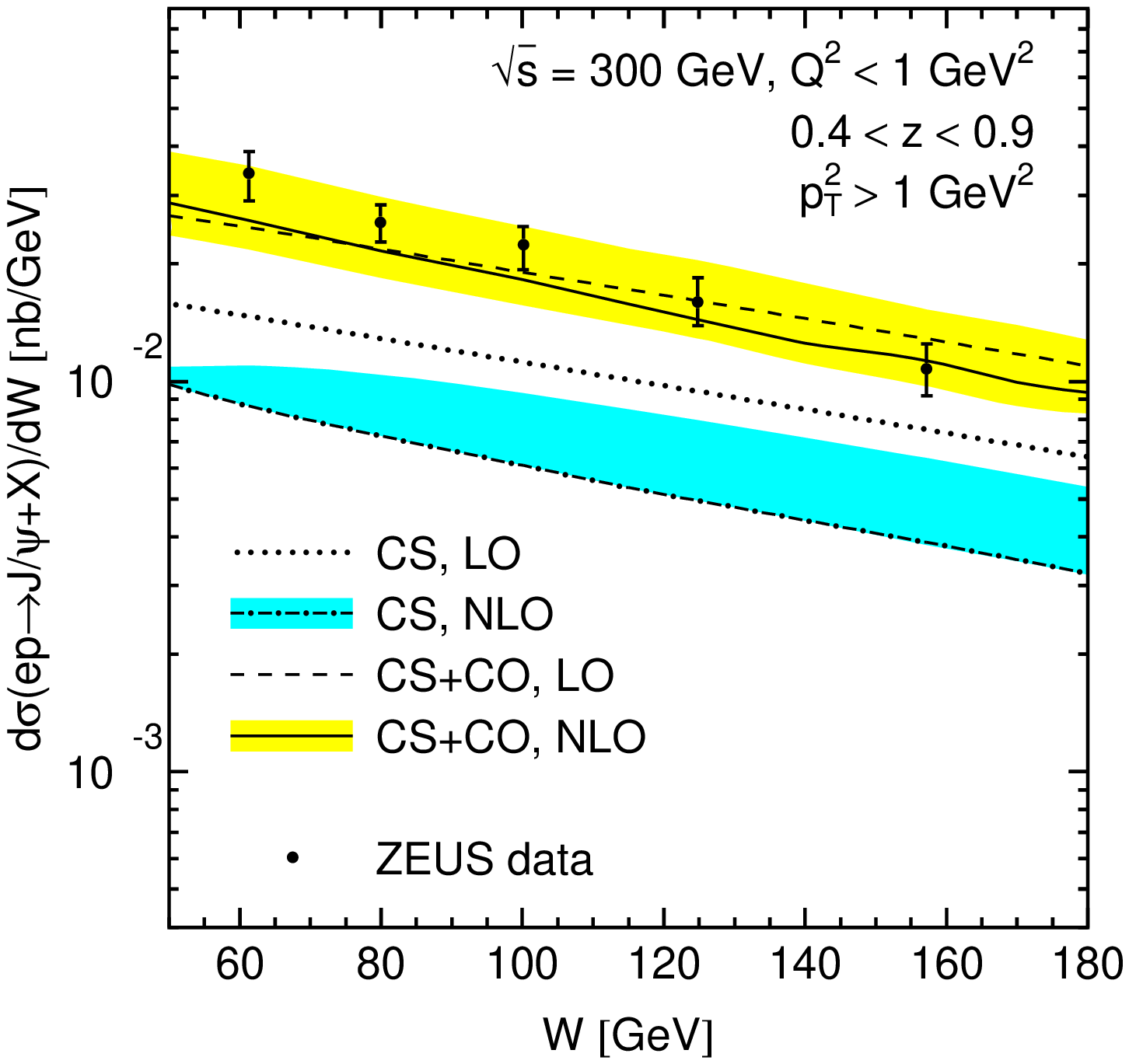}
\includegraphics[width=0.225\textwidth]{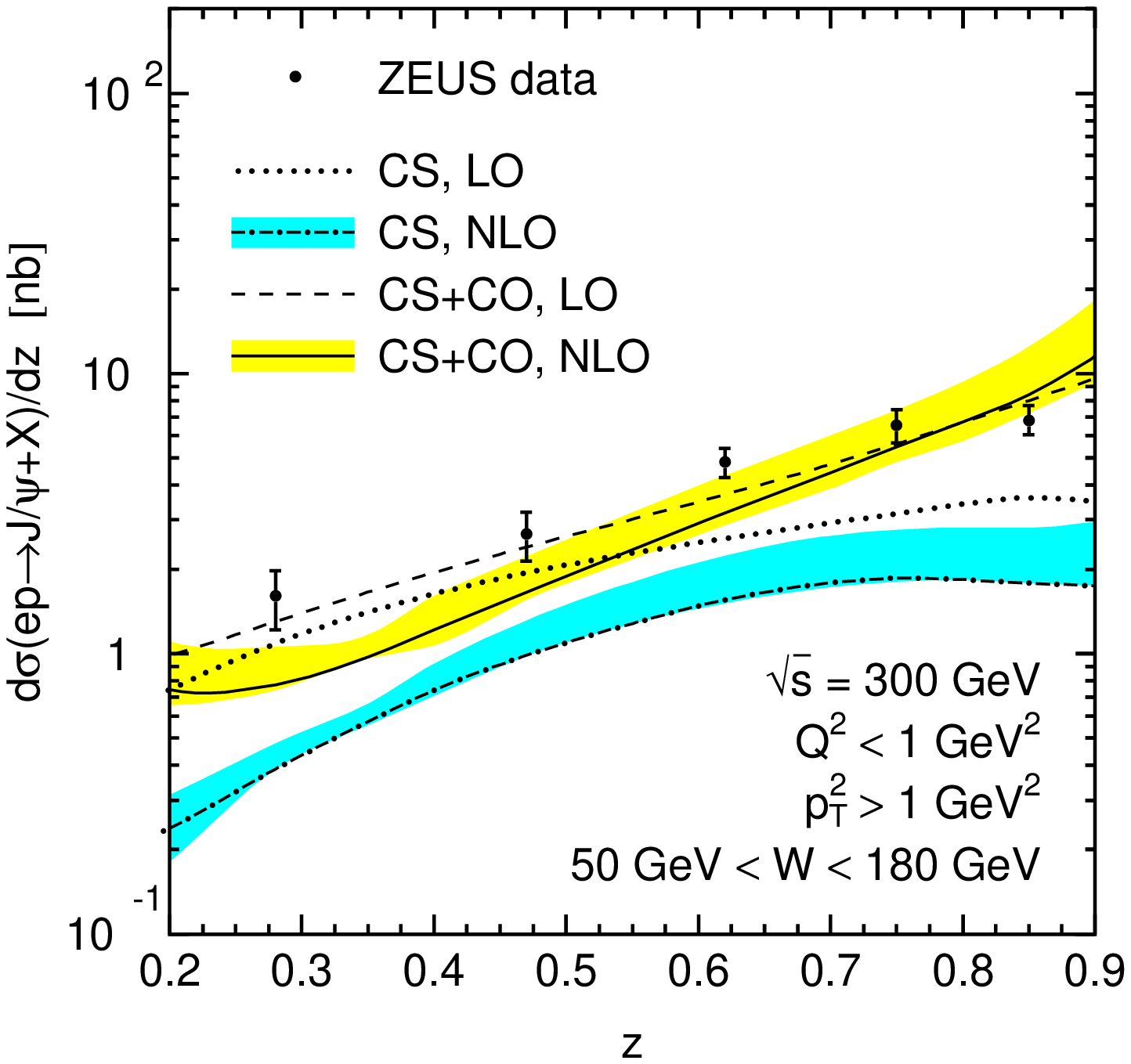}
\includegraphics[width=0.225\textwidth]{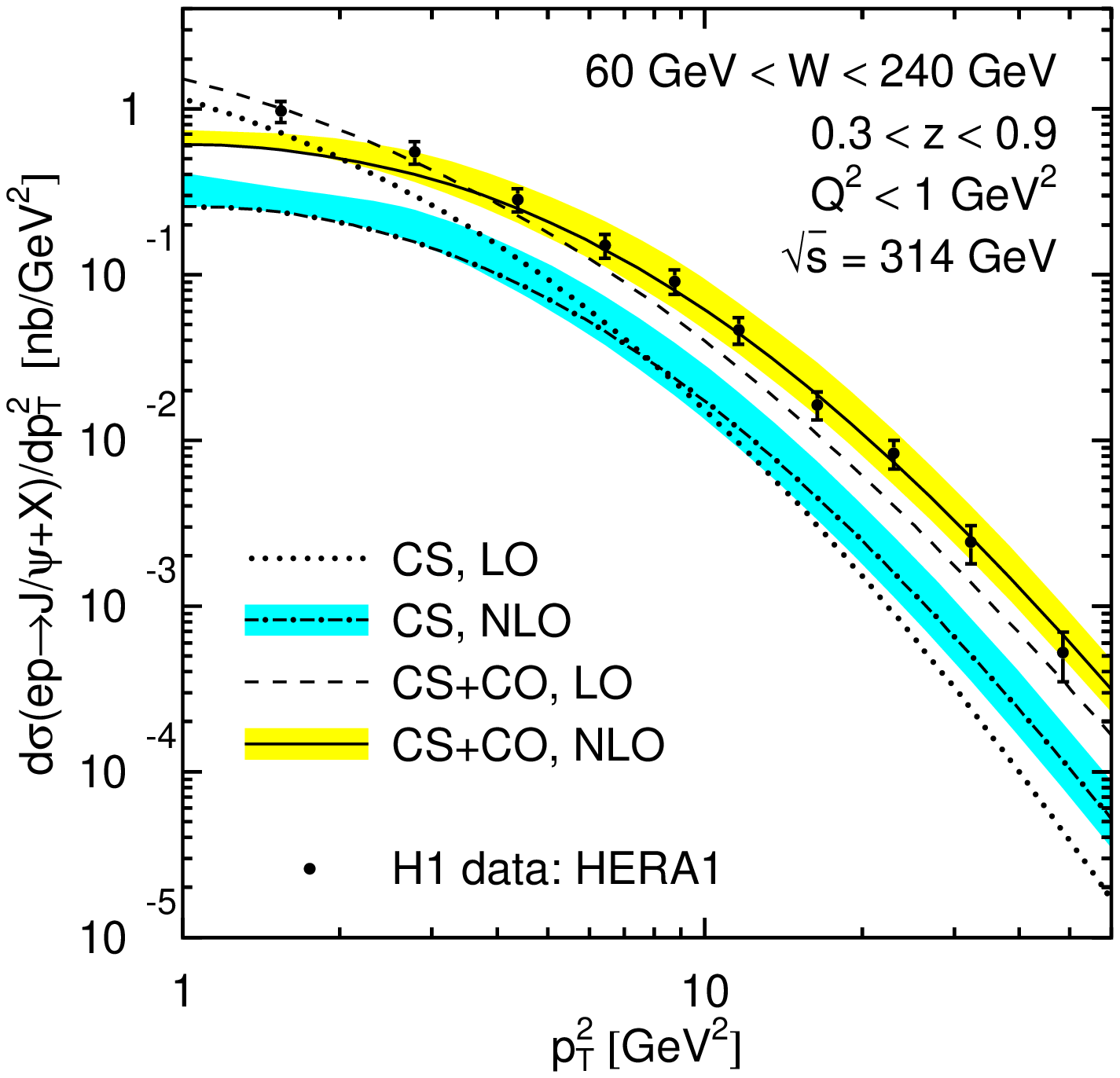}
\includegraphics[width=0.225\textwidth]{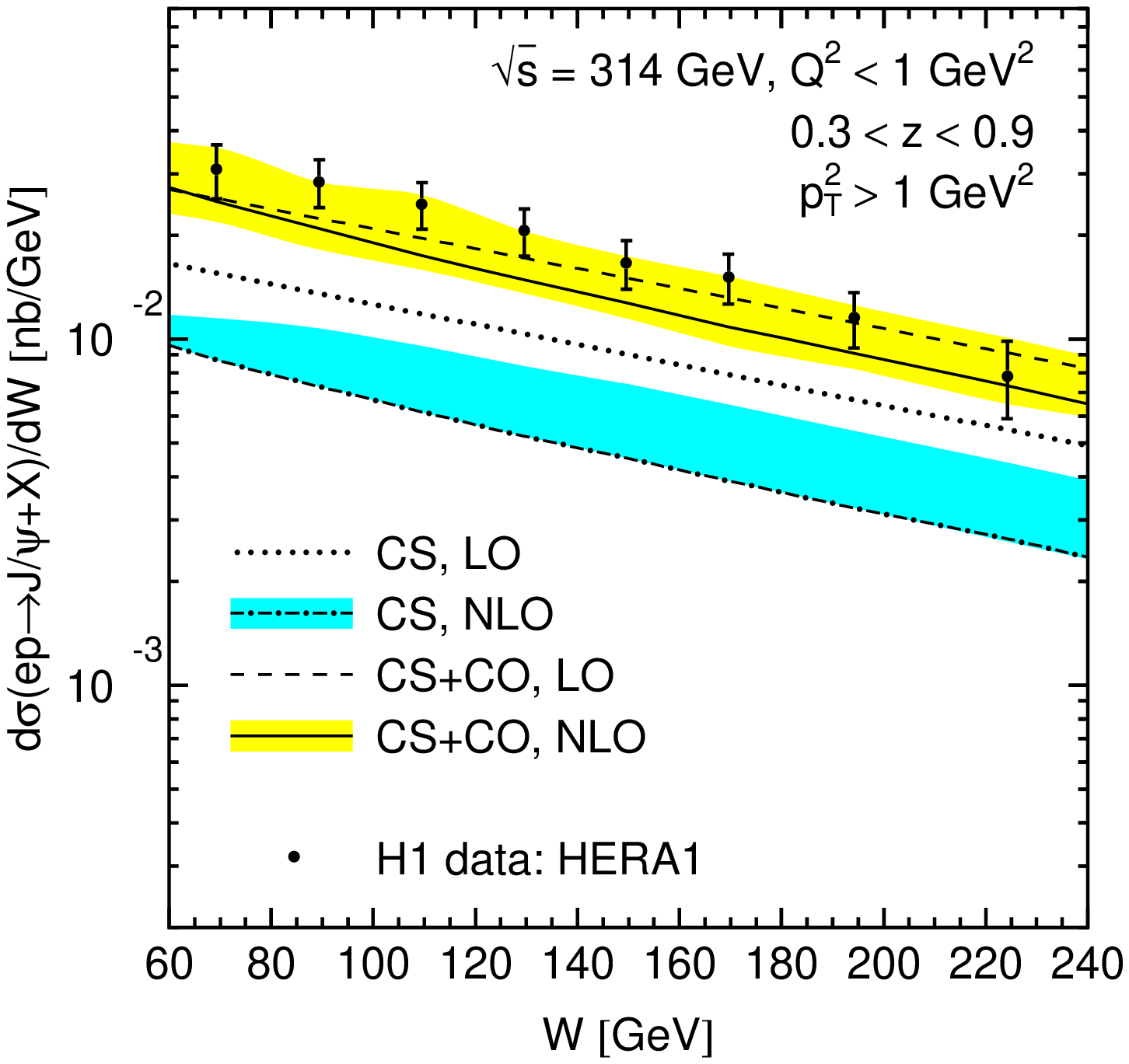}

\vspace{2pt}
\includegraphics[width=0.225\textwidth]{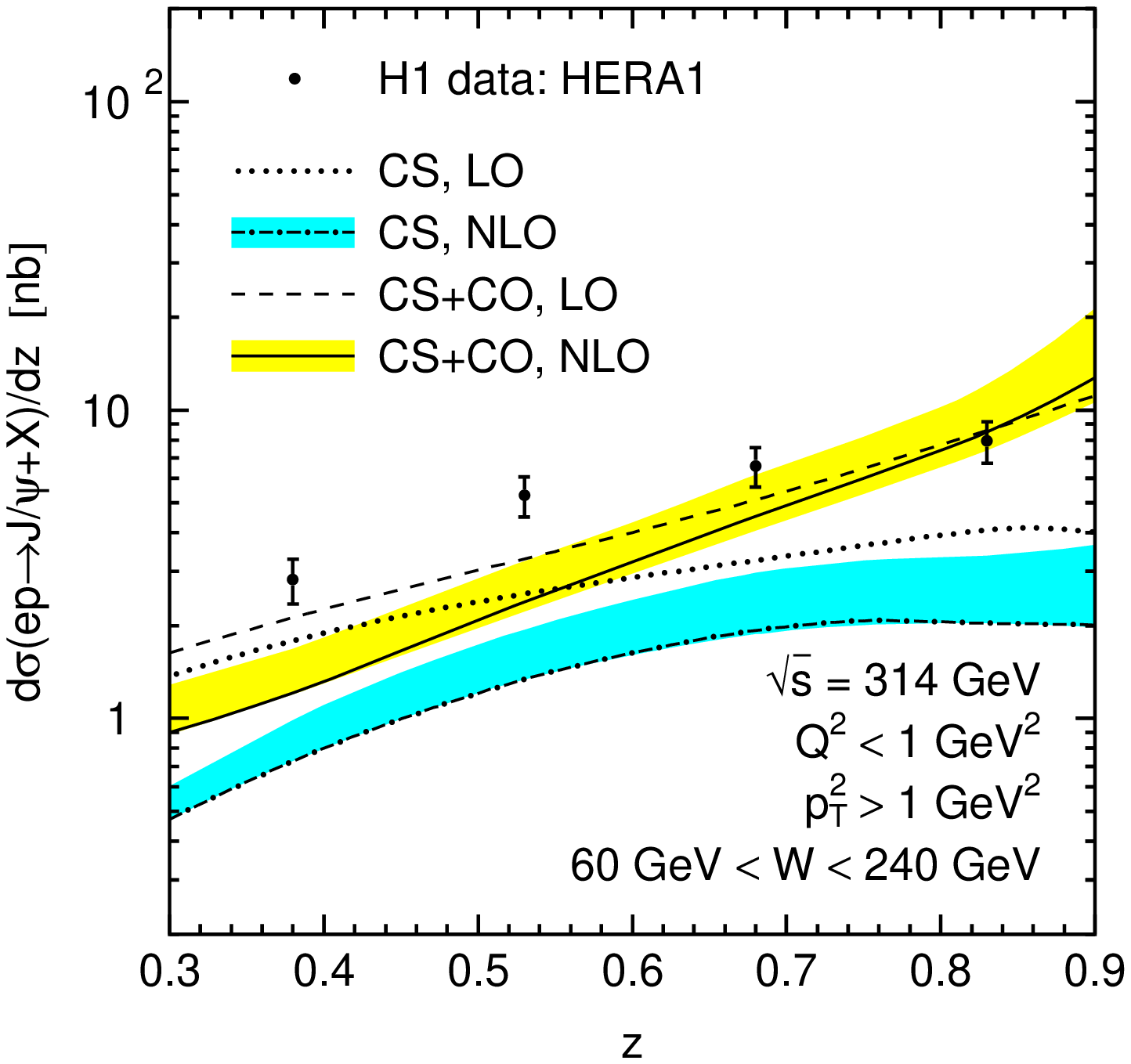}
\includegraphics[width=0.225\textwidth]{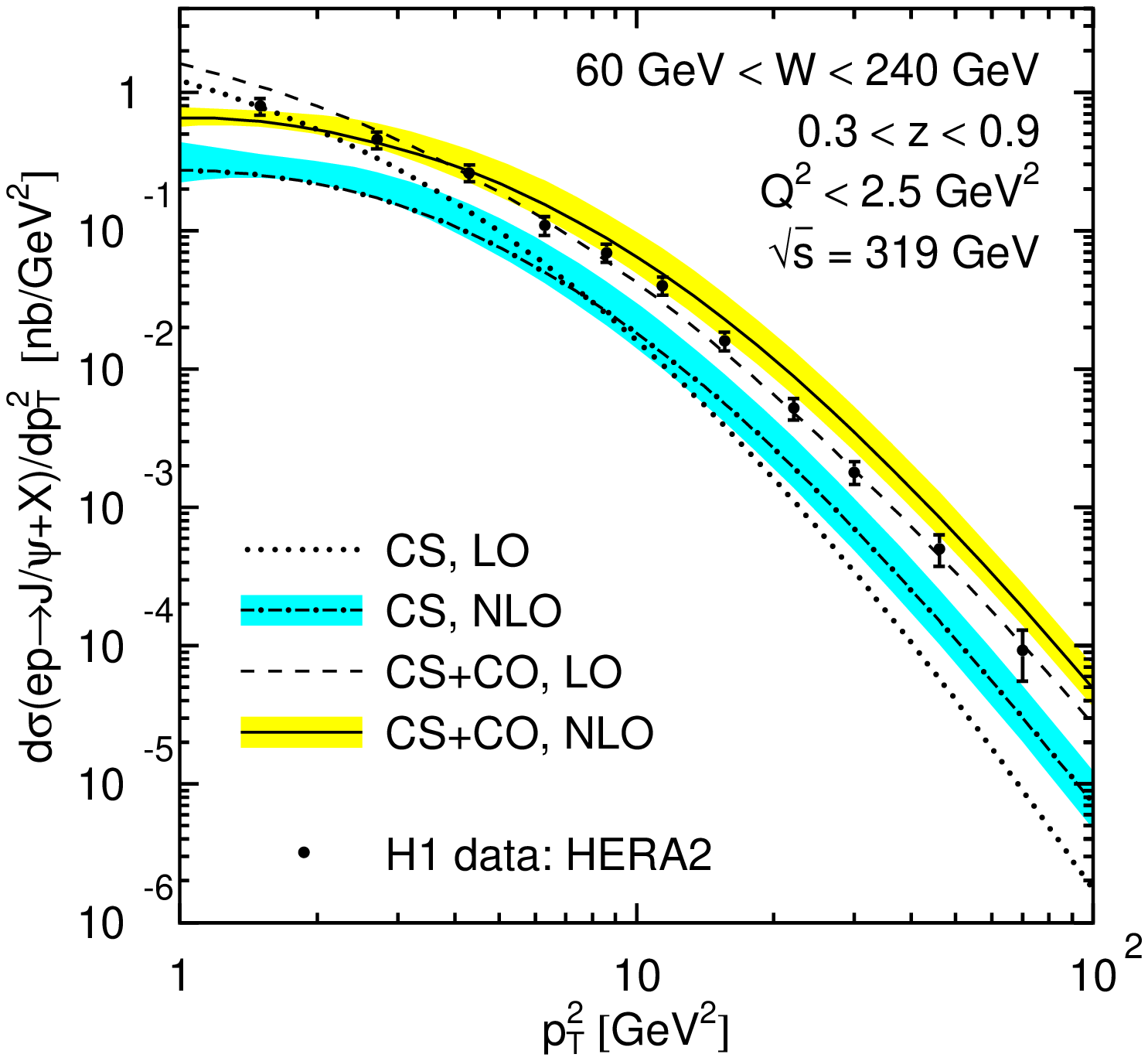}
\includegraphics[width=0.225\textwidth]{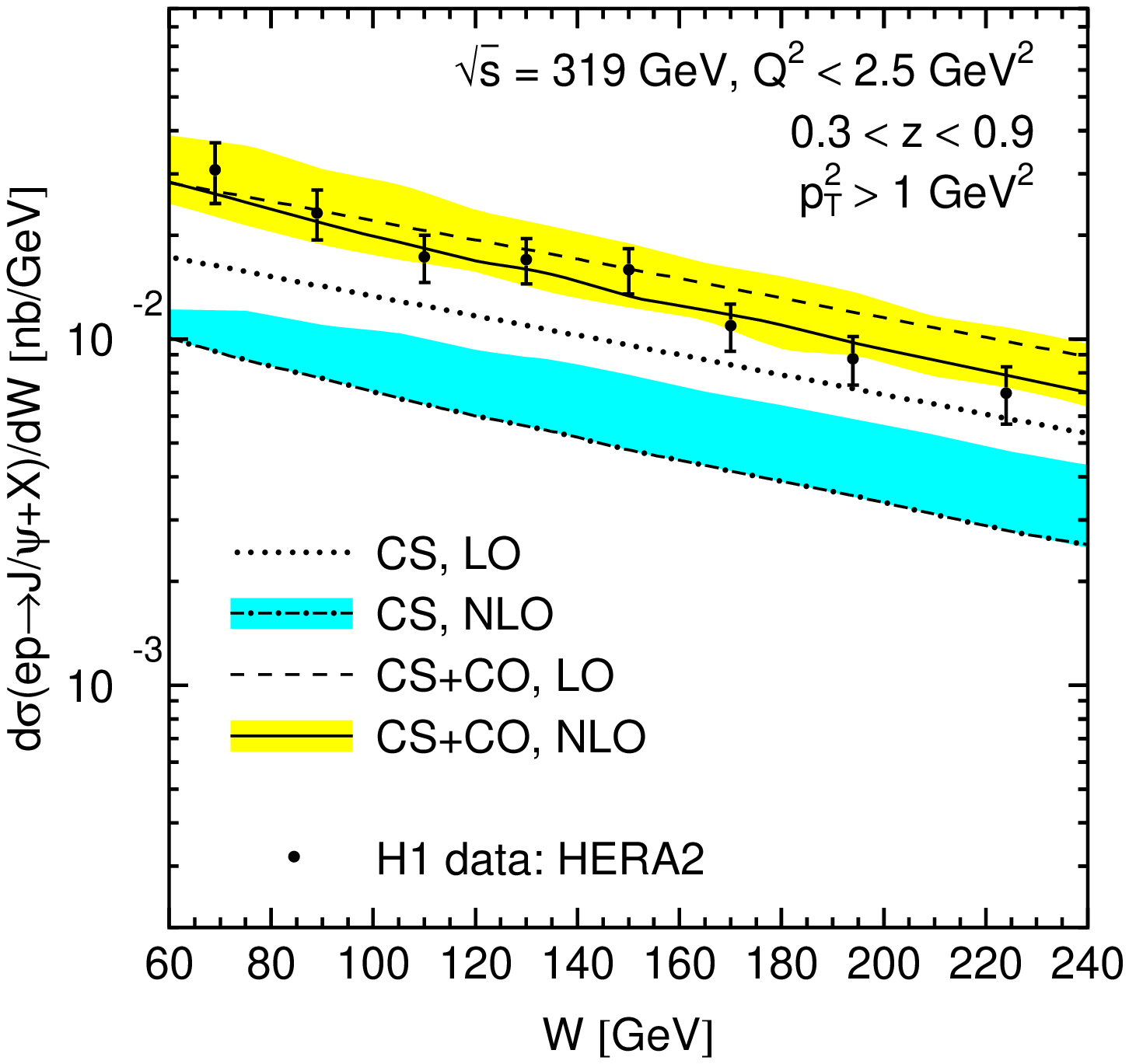}
\includegraphics[width=0.225\textwidth]{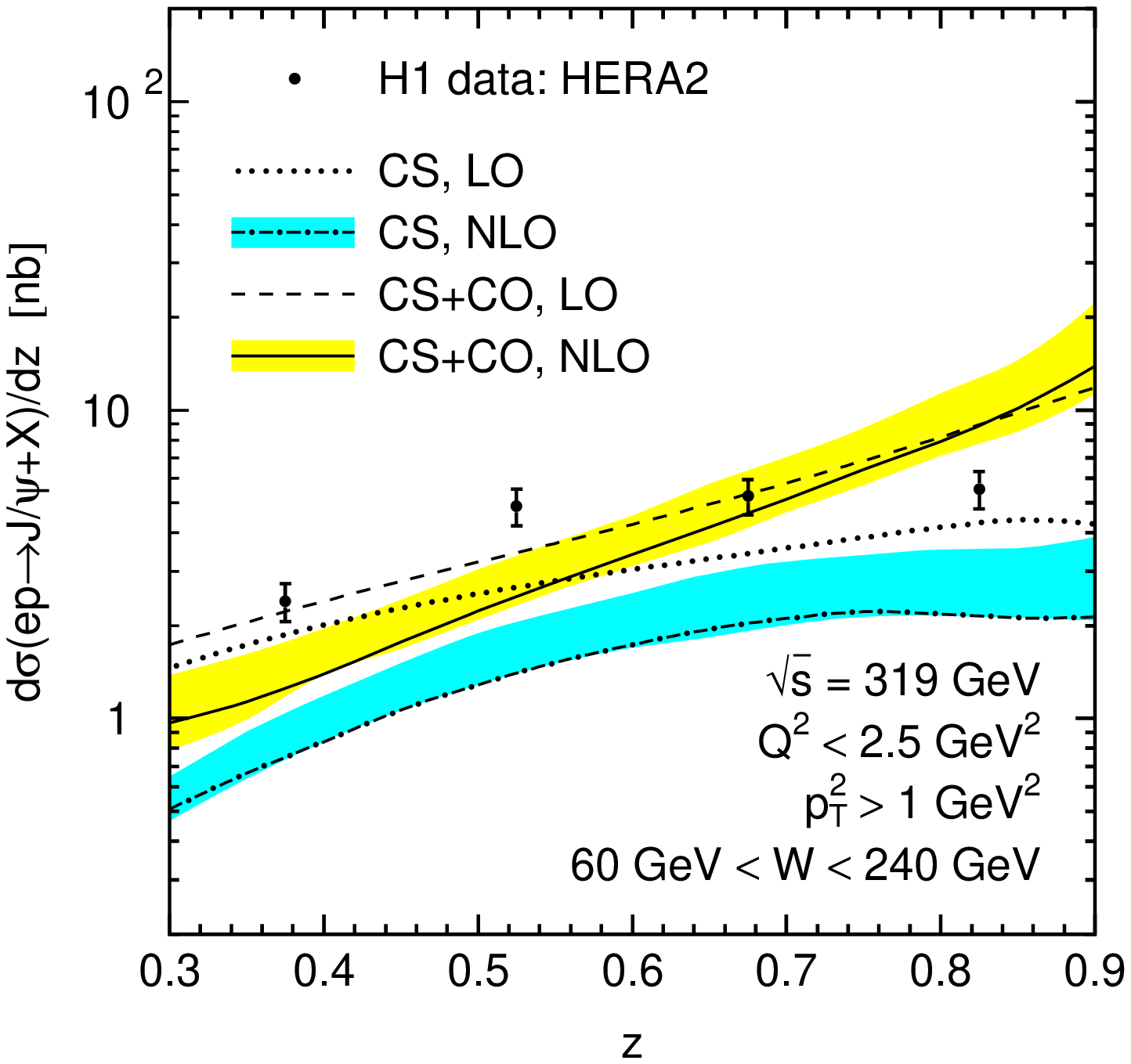}
\caption{\label{fig:fitgraphs}%
NLO NRQCD fit compared to RHIC \cite{Adare:2009js}, Tevatron
\cite{Acosta:2004yw,Abe:1997jz}, LHC
\cite{Khachatryan:2010yr,ATLASdata,ALICEdata,LHCbdata}, and HERA
\cite{Adloff:2002ex,Aaron:2010gz,Chekanov:2002at} data.}
\end{figure*}
We exclude from our fit all data points of photoproduction and two-photon
scattering with $p_T<1$~GeV and of hadroproduction with $p_T<3$~GeV, which
cannot be successfully described by our fixed-order calculations as expected.
This leaves a total of 194 data points.
The fit results for the CO LDMEs obtained at NLO in NRQCD with default scale
choices are collected in Table~\ref{tab:fit}.
They depend only feebly on the precise locations of the $p_T$ cuts.
In the following, we use the values of Table~\ref{tab:fit} throughout.

\begin{figure*}
\begin{tabular}{cccc}
\includegraphics[width=0.225\textwidth]{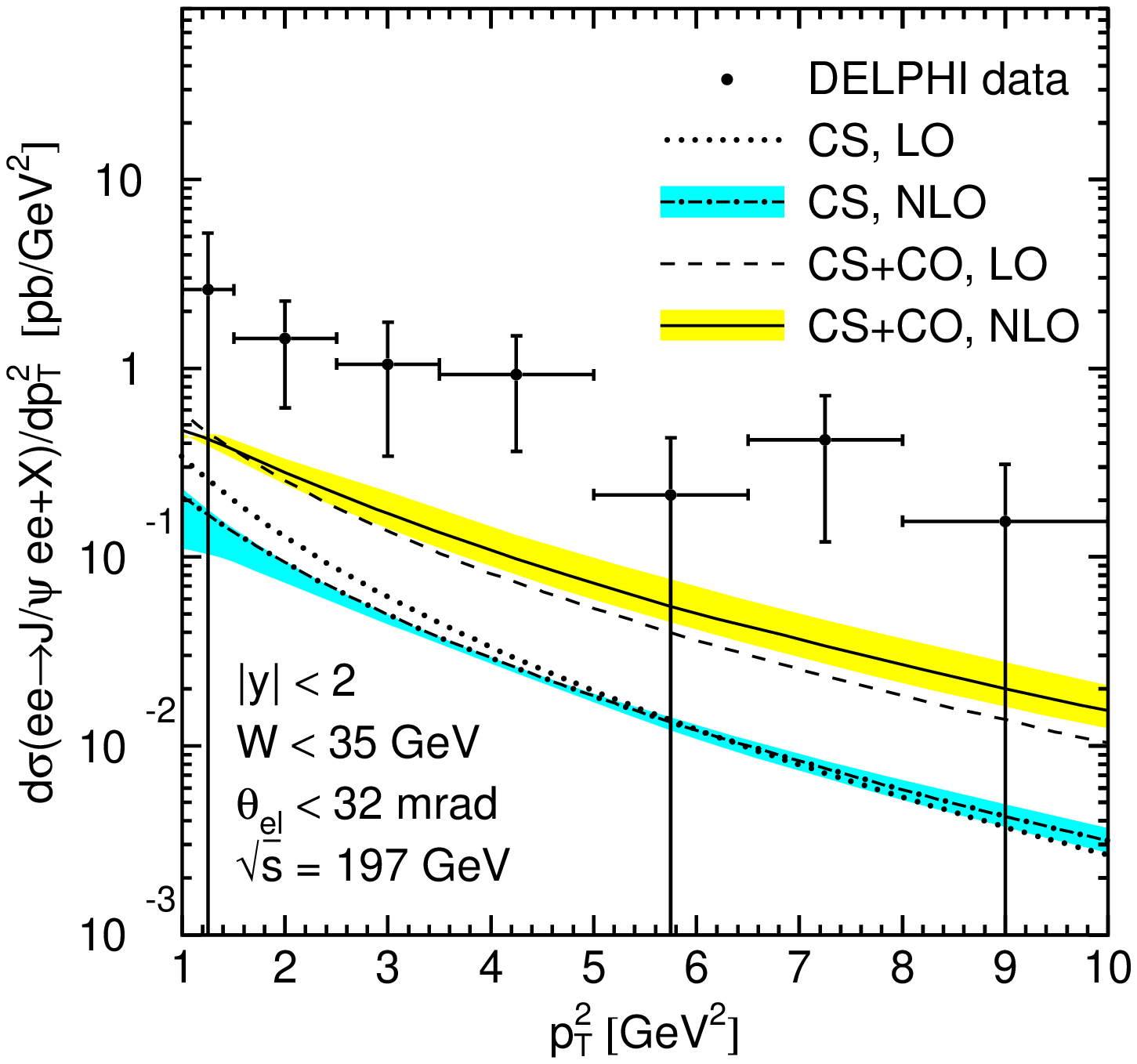}
&
\includegraphics[width=0.225\textwidth]{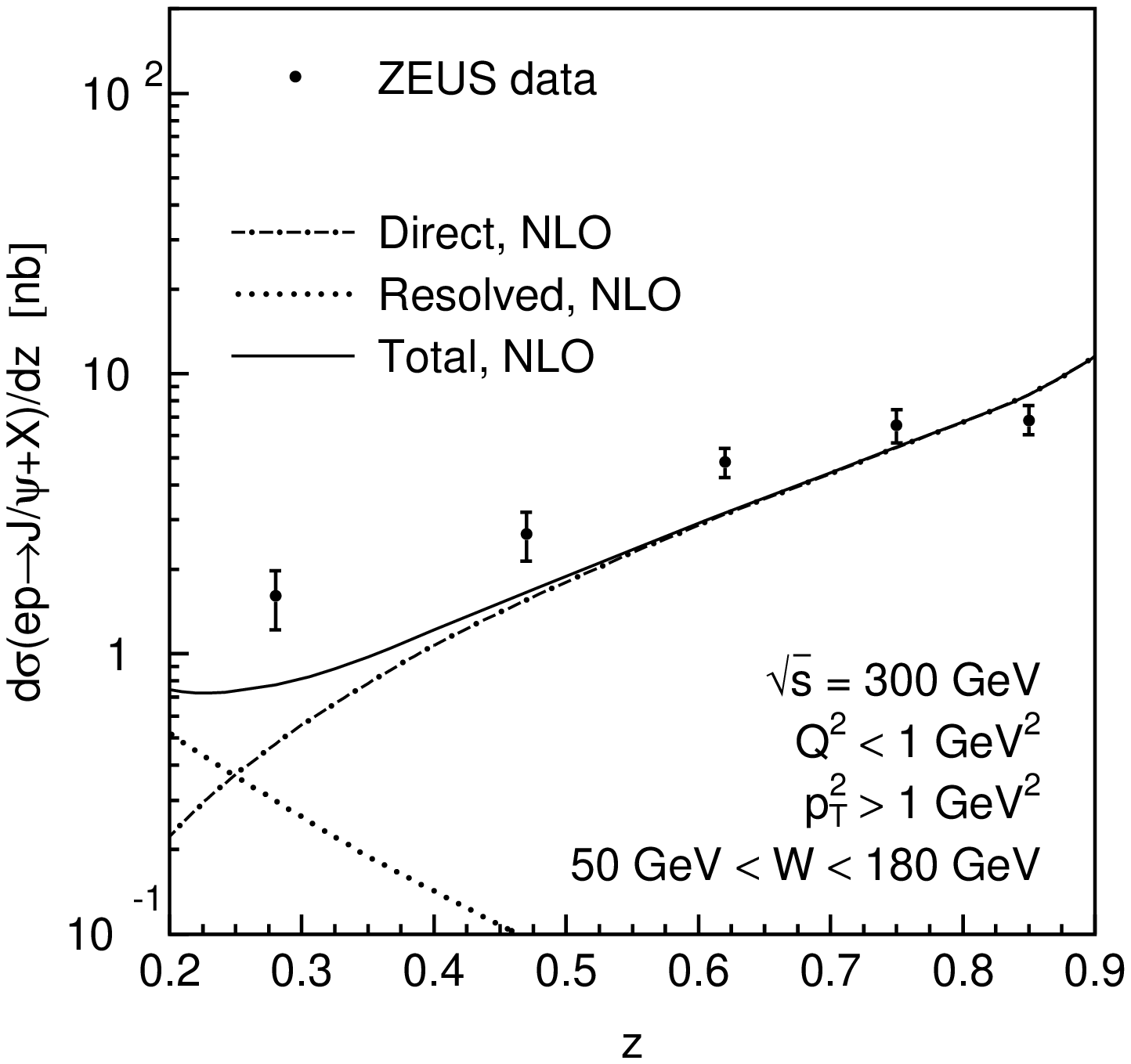}
&
\includegraphics[width=0.225\textwidth]{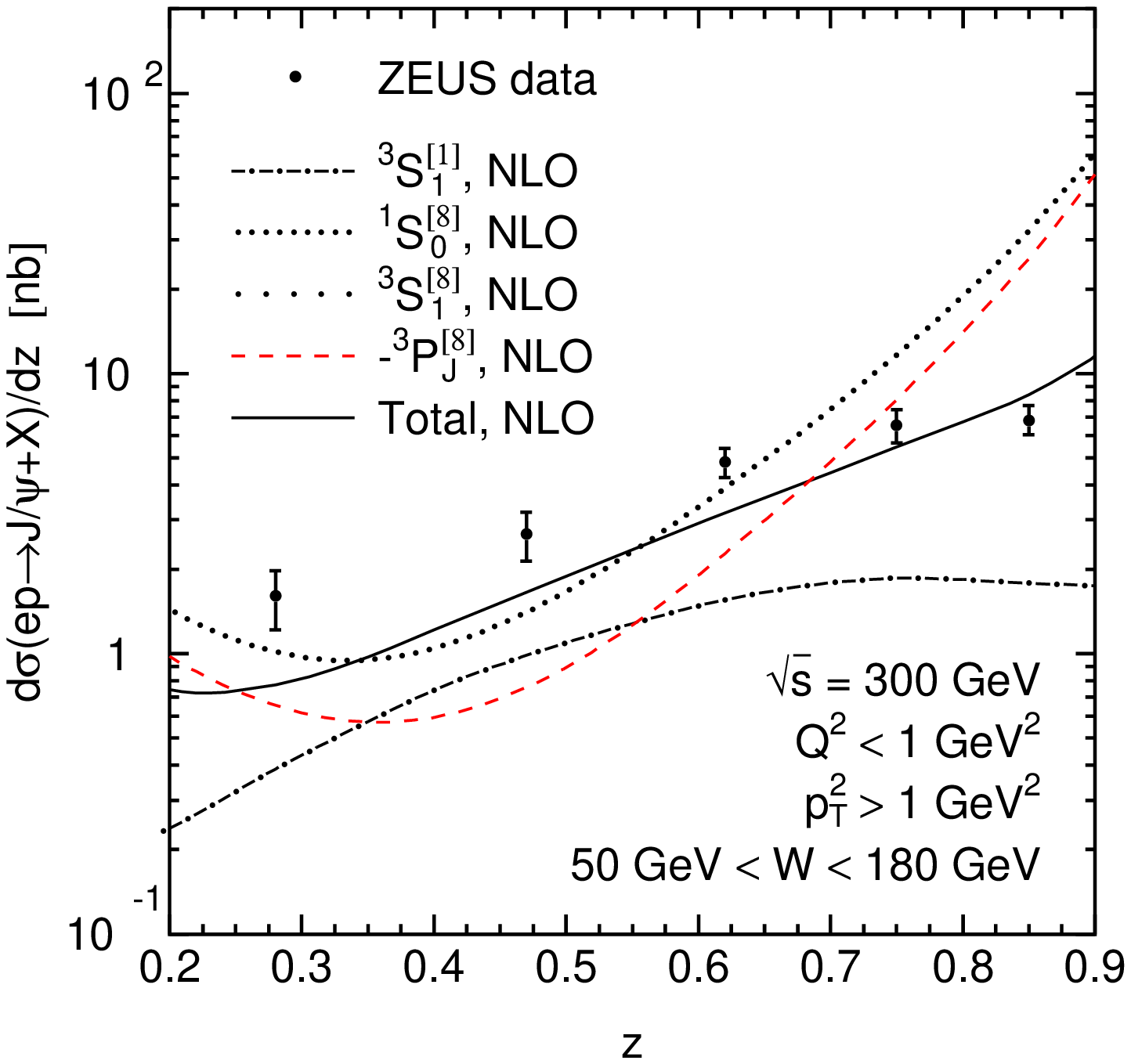}
&
\includegraphics[width=0.225\textwidth]{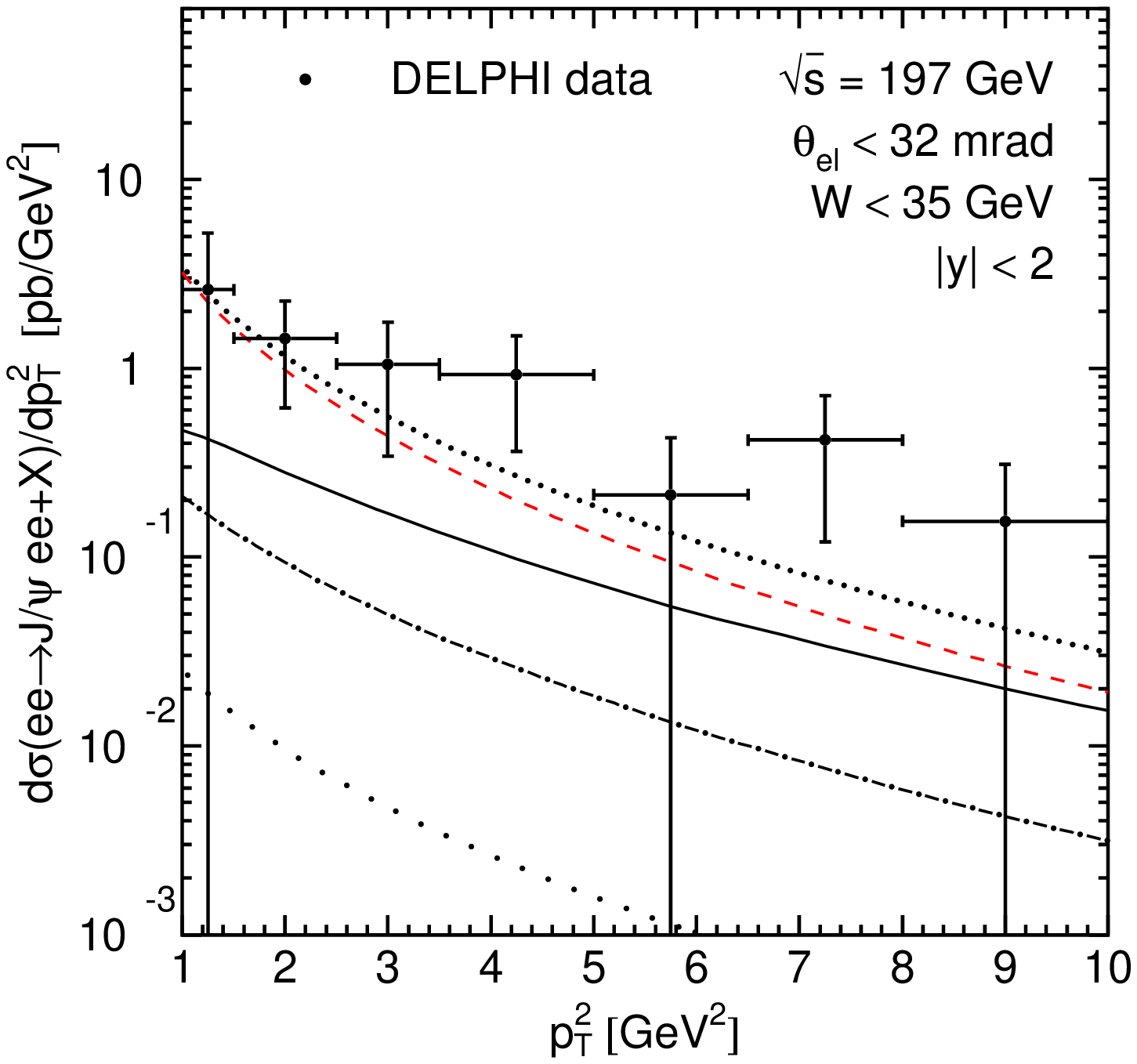}\\
\vspace{-0.85cm} & & & \\
(a) \hspace{2.9cm} & (b) \hspace{2.9cm} & (c) \hspace{2.9cm} & (d)
\hspace{2.9cm} \\
\end{tabular}
\caption{\label{fig:states}
(a) NLO NRQCD fit compared to LEP~II \cite{Abdallah:2003du} data and (d) its
decomposition to $c\overline{c}[n]$ channels.
Decomposition of $z$ distribution at HERA~I \cite{Chekanov:2002at} (b) to
contributions due to direct and resolved photoproduction and (c) to 
$c\overline{c}[n]$ channels.} 
\end{figure*}
In Figs.~\ref{fig:fitgraphs} and \ref{fig:states}(a), all data sets fitted to,
except the single data point from Belle \cite{:2009nj}, are compared with our
default NLO NRQCD results (solid lines).
For comparison, also the default predictions at LO (dashed lines) as well as
those of the CSM at NLO (dot-dashed lines) and LO (dotted lines) are shown.
The yellow and blue (shaded) bands indicate the theoretical errors on the
NLO NRQCD and CSM predictions.
We observe from Figs.~\ref{fig:fitgraphs} that the experimental data are nicely
described by NLO NRQCD, being almost exclusively contained within its error
bands, while they overshoot the NLO CSM predictions typically by 1--2 orders of
magnitude for hadroproduction and a factor of 3--5 for photoproduction.
The description of the $z$ distributions in photoproduction by NLO NRQCD
significantly benefits from two features, rendering it considerably more
favorable than in Refs.~\cite{Butenschoen:2009zy,Butenschoen:2010rq}.
On the one hand, as conjectured in
Refs.~\cite{Butenschoen:2009zy,Butenschoen:2010rq}, resolved photoproduction
usefully enhances the cross section in the low-$z$ range, being dominant for
$z\alt0.25$, as is evident from Fig.~\ref{fig:states}(b).
On the other hand, owing to the negative value of
$\langle{\cal O}^{J/\psi}(^3\!P_0^{[8]})\rangle$ in Table~\ref{tab:fit}, the
$^1\!S_0^{[8]}$ and $^3\!P_J^{[8]}$ contributions interfere destructively thus
attenuating the familiar rise in cross section in the limit $z\to1$, as may be
seen from Fig.~\ref{fig:states}(c).
As for the $p_T^2$ and $W$ distributions in photoproduction, the cut $z>0.3$
(0.4) applied by H1 (ZEUS) greatly suppresses resolved photoproduction, to the
level of 1\%.
In contrast to the LO analysis of Ref.~\cite{Klasen:2001cu}, the DELPHI
\cite{Abdallah:2003du} data tend to systematically overshoot the NLO NRQCD
result, albeit the deviation is by no means significant in view of the
sizeable experimental errors.
As is evident from Fig.~\ref{fig:states}(d), this may be attributed to the
destructive interference of the $^1\!S_0^{[8]}$ and $^3\!P_J^{[8]}$
contributions mentioned above, which is a genuine NLO phenomenon.
We have to bear in mind, however, that the DELPHI measurement comprises only 16
events with $p_T>1$~GeV and has not been confirmed by any of the other three
LEP~II experiments.
In two-photon scattering at LEP~II, the single-resolved contribution vastly
dominates over the direct and double-resolved ones, as was already observed
for the LO case in Ref.~\cite{Klasen:2001cu}.
The Belle measurement, $\sigma(e^+e^-\to J/\psi+X)=(0.43\pm0.13)$~pb, is
compatible both with the NLO NRQCD and CSM results,
$(0.70{+0.35\atop-0.17})$~pb and $(0.24{+0.20\atop-0.09})$~pb, respectively;
at LO, where $X=g$, we are dealing with a pure CO process, with a total cross
section of 0.23~pb.
The overall goodness $\chi_{\rm d.o.f.}^2=857/194=4.42$ of our NLO NRQCD fit,
which we quote for completeness, is of limited informative value, since the
theoretical uncertainties exceed most of the experimental errors.

Our theoretical predictions refer to direct $J/\psi$ production, as the data
from Tevatron~I \cite{Abe:1997jz} do, while the data from KEKB \cite{:2009nj},
Tevatron~II \cite{Acosta:2004yw}, and LHC
\cite{Khachatryan:2010yr,ATLASdata,ALICEdata,LHCbdata} comprise prompt events
and those from LEP~II \cite{Abdallah:2003du}, HERA
\cite{Adloff:2002ex,Aaron:2010gz,Chekanov:2002at}, and RHIC \cite{Adare:2009js}
even non-prompt ones.
However, the resulting error is small against our theoretical uncertainties and
has no effect on our conclusions.
In fact, the fraction of $J/\psi$ events originating from the feed-down of
heavier charmonia only amounts to about 36\% \cite{Abe:1997jz} for
hadroproduction, 15\% \cite{Aaron:2010gz} for photoproduction at HERA,
9\% for two-photon scattering at LEP~II \cite{Klasen:2004tz}, and 26\% for
$e^+e^-$ annihilation at KEKB \cite{Ma:2008gq}, and the
fraction of $J/\psi$ events from $B$ decays is negligible RHIC, HERA
\cite{Aaron:2010gz}, and LEP~II \cite{Klasen:2004tz} energies.
Refitting the data with the estimated feed-down contribution subtracted yields
$\langle {\cal O}^{J/\psi}(^1\!S_0^{[8]}) \rangle=
(3.04\pm0.35)\times10^{-2}$~GeV$^3$,
$\langle {\cal O}^{J/\psi}(^3\!S_1^{[8]}) \rangle=
(1.68\pm0.46)\times10^{-3}$~GeV$^3$, and
$\langle {\cal O}^{J/\psi}(^3\!P_0^{[8]}) \rangle=
(-9.08\pm1.61)\times10^{-3}$~GeV$^5$ with a slightly reduced value
$\chi_{\rm d.o.f.}^2=725/194=3.74$.

In conclusion, we performed a NLO NRQCD analysis of all available high-quality
data of inclusive unpolarized $J/\psi$ production, from KEKB \cite{:2009nj},
LEP~II \cite{Abdallah:2003du}, RHIC \cite{Adare:2009js}, HERA~I
\cite{Adloff:2002ex,Chekanov:2002at} and II \cite{Aaron:2010gz}, Tevatron~I
\cite{Abe:1997jz} and II \cite{Acosta:2004yw}, and the LHC
\cite{Khachatryan:2010yr,ATLASdata,ALICEdata,LHCbdata}, comprising a total of
194 data points from 26 data sets.
The fit values of the CO LDMEs in Table~\ref{tab:fit} agree with our previous
ones \cite{Butenschoen:2010rq}, extracted just from the $p_T$ distributions of
Refs.~\cite{Adloff:2002ex,Aaron:2010gz,Acosta:2004yw}, within the errors of the
latter, but the new errors are about 40\% smaller.
In compliance with the velocity scaling rules of NRQCD \cite{Bodwin:1994jh},
these values are approximately of order ${\cal O}(v^4)$ relative to
$\langle {\cal O}^{J/\psi}(^3\!S_1^{[1]}) \rangle$.
This manifestly consolidates the verification of NRQCD factorizatization for
charmonium and provides rigorous evidence for LDME universality and the
existence of CO processes in nature.

This work was supported in part by BMBF Grant No.\ 05H09GUE, DFG Grant
No.\ KN~365/6--1, and HGF Grant No.\ HA~101.

\end{document}